\documentclass[usenatbib]{mn2e}
\usepackage{graphicx,tabularx,amsmath,amssymb,upgreek,wasysym,mathtools,multirow,siunitx}
\numberwithin{equation}{section}

\newcommand{\subD}{_{_{\rm D}}}
\newcommand{\subO}{_{_{\rm O}}}
\newcommand{\pp}{{\sc ppmap}$\;$} 
\newcommand{\ppp}{{\sc ppmap}}

\def\apjl{{\em ApJ}}
\def\apjs{{\em ApJS}}
\def\aap{{\em A\&A}}

\title[The dust in M31]{The dust in M31}
\author[A. P. Whitworth et al.]
{A. P. Whitworth,$^{1}$\thanks{E-mail: ant@astro.cf.ac.uk} 
 K. A. Marsh,$^{2}$ 
 P. J. Cigan,$^{1}$ 
 J. J. Dalcanton,$^{3}$ 
 \newauthor
 M. W. L. Smith,$^{1}$ 
 H. L. Gomez,$^{1}$ 
 O. Lomax,$^{4}$ 
 M. J. Griffin$^{1}$ 
 and S. A. Eales$^{1}$ \\
$^{1}$School of Physics and Astronomy, Cardiff University, Cardiff CF24 3AA, Wales, UK \\
$^{2}$IPAC, CalTech, 1200E California Boulevard Pasadena, CA91125, USA \\
$^{3}$Department of Astronomy, University of Washington, Box 351580, Seattle, WA98195, USA \\
$^{4}$ESTEC, Keplerlaan 1, 2201 AZ Noordwijk, Netherlands}
\begin{document}
\pagerange{\pageref{firstpage}--\pageref{lastpage}} \pubyear{2013}
\maketitle
\label{firstpage}

\begin{abstract}
We have analysed {\it Herschel} observations of M31, using the \pp procedure. The resolution of \pp images is sufficient ($\sim 31\,{\rm pc}$ on M31) that we can analyse far-IR dust emission on the scale of Giant Molecular Clouds. By comparing \pp estimates of the far-IR emission optical depth at $300\,\mu{\rm m}\,(\tau_{_{300}})$, and the near-IR extinction optical depth at $1.1\,\mu{\rm m}\,(\tau_{_{1.1}})$ obtained from the reddening of RGB stars, we show that the ratio ${\cal R}^{\mbox{\tiny obs.}}_\tau\equiv\tau_{_{1.1}}/\tau_{_{300}}$ falls in the range $500\la{\cal R}^{\mbox{\tiny obs.}}_\tau\la 1500$. Such low values are incompatible with many commonly used theoretical dust models, which predict values of ${\cal R}^{\mbox{\tiny model}}_\kappa\equiv\kappa_{_{1.1}}/\kappa_{_{300}}$ (where $\kappa$ is the dust opacity coefficient) in the range $2500\la{\cal R}^{\mbox{\tiny model}}_\kappa\la 4000$. That is, unless a large fraction, $\ga 60\%$, of the dust emitting at $300\,\mu{\rm m}$ is in such compact sources that they are unlikely to intercept the lines of sight to a distributed population like RGB stars. This is not a new result: variants obtained using different observations and/or different wavelengths have already been reported by other studies. We present two analytic arguments for why it is unlikely that $\ga 60\%$ of the emitting dust is in sufficiently compact sources. Therefore it may be necessary to explore the possibility that the discrepancy between observed values of ${\cal R}^{\mbox{\tiny obs.}}_\tau$ and theoretical values of ${\cal R}^{\mbox{\tiny model}}_\kappa$ is due to limitations in existing dust models. \pp also allows us to derive optical-depth weighted mean values for the emissivity index, $\beta\equiv -d\ln(\kappa_{_\lambda})/d\ln(\lambda)$, and the dust temperature, $T$, denoted ${\bar\beta}$ and ${\bar T}$. We show that, in M31, ${\cal R}^{\mbox{\tiny obs.}}_\tau$ is anti-correlated with ${\bar\beta}$ according to ${\cal R}^{\mbox{\tiny obs.}}_\tau\simeq 2042(\pm 24)-557(\pm 10){\bar\beta}$. If confirmed, this provides a challenging constraint on the nature of interstellar dust in M31.
\end{abstract}

\begin{keywords}
ISM: dust, extinction -- submillimetre: galaxies -- galaxies: Local Group, structure, ISM
\end{keywords}

\section{Introduction}\label{SEC:Intro}

\subsection{Preamble}

Much of the solid material in the Universe is in the form of interstellar dust \citep[e.g.][]{DraineBT2003}. This dust is the material which forms planets; it is the material which plays a vital role in cooling gas as it condenses into new stars; and it is the material which seriously compromises our view of the hot objects in the Universe, by absorbing a significant fraction of their light, and then re-emitting it at far-infrared wavelengths. Despite its importance, our understanding of the nature of interstellar dust is limited.

\subsection{The galaxies of the Local Group}

The Local Group contains two major disc galaxies: the Milky Way and M31. They have comparable masses and extents, and are separated by $\sim\!0.78\,{\rm Mpc}$ \citep{Richetal2005}. Because we live in it, our view of the Milky Way is detailed, but confused, due to the superposition of sources at different distances, distance uncertainties, and dust extinction. Our view of M31 is less detailed, but the large-scale layout and dynamics of its disc are relatively clear. The molecular clouds and star formation in M31 are concentrated in three rings, at radii of $\sim\!6\,{\rm kpc}$, $\sim\!11\,{\rm kpc}$ and $\sim\!15\,{\rm kpc}$; the middle ring is the most massive, and has the highest star formation rate \citep{Lewietal2015}. Although structural details of M31's disc differ from the Milky Way,  there is no evidence that the dust in M31 is markedly different from that in the Milky Way, a disc galaxy of comparable size, age and environment \citep{Clayetal2015}. However, we should be mindful that in more distant galaxies -- of different type, size, age and/or environment -- the properties of dust might be significantly different.

\subsection{Analysing {\it Herschel} maps with {\sc ppmap}}

We have used the \pp procedure \citep{Marsetal2015} to analyse {\it Herschel} {\sc pacs} images from Krause et al. \citep[unpublished; see][]{Grovetal2012}, with mean wavelengths (and mean beam sizes) of $70\,\mu{\rm m}$ ({\sc fwhm}$=8.5''$),  $100\,\mu{\rm m}$ ({\sc fwhm}$=12.5''$) and $160\,\mu{\rm m}$ ({\sc fwhm}$=13.3''$); and {\it Herschel} {\sc spire} images from \citet{Fritetal2012}, with mean wavelengths (and mean beam sizes) of $250\,\mu{\rm m}$ ({\sc fwhm}$=18.2''$), $350\,\mu{\rm m}$ ({\sc fwhm}$=24.5''$) and $500\,\mu{\rm m}$ ({\sc fwhm}$=36.0''$). For the {\sc pacs} observations, we use the azimuthally averaged PSFs from \citet{Pogletal2010} adjusted for blurring induced by the $20''\,{\rm s}^{-1}$ scanning speed. For the {\sc spire} observations, we use the azimuthally averaged PSFs from \citet{Grifetal2010}. No beamshape corrections are made for the spectral shape. It would be straightforward to include such corrections in the {\sc ppmap} procedure, but in practice they are not significant. We do correct for the spectral shape of the bandpass.

By abandoning the restrictive assumptions underlying the standard procedure for analysing far-infrared dust emission, \pp not only produces separate images of the optical depth of emitting dust of different types, and in different temperature intervals, it also achieves much higher spatial resolution ($\sim 31\,{\rm pc}$). Consequently we can evaluate the total emission optical depth more accurately and at higher resolution. We can also constrain which types of dust and which temperature intervals make the major contributions to the total emission optical depth. 

By comparing images of the far-infrared dust-emission optical depth at $300\,\mu{\rm m}$ $\,(\tau_{_{300}})$ with images of the near-infrared dust-extinction optical depth at $1.1\,\mu{\rm m}$ $\,(\tau_{_{1.1}})$ \citep{Dalcetal2015}, we can compute the ratio ${\cal R}^{\mbox{\tiny obs.}}_\tau=\tau_{_{1.1}}/\tau_{_{300}}$ in 28726 individual $15\rm{pc}\times 15\rm{pc}$ pixels. We can also compute the optical-depth-weighted mean of the far-IR emissivity index, $\beta= -\,\left.d\ln\left(\tau_{_\lambda}\right)/d\ln(\lambda)\right|_{_{\rm \lambda =300\mu m}}$, on the line of sight through each pixel, and similarly the optical-depth-weighted mean of the dust temperature, $T$, on the line of sight through each pixel.

\subsection{Plan of paper}

This paper has to do with the statistics of the above quantities (${\cal R}^{\mbox{\tiny obs.}}_\tau,\beta,T$), and what they might be telling us about the properties of interstellar dust. Section \ref{SEC:StanProc} outlines the standard procedure used to analyse far-infrared observations of dust emission, and the limitations of this procedure. Section \ref{SEC:PPMAP} outlines the \pp procedure, its advantages and limitations. Section \ref{SEC:Results} describes and illustrates the results of applying \pp to {\it Herschel} observations of M31. Section \ref{SEC:Correlations} discusses the observed correlations between derived dust properties. Section \ref{SEC:CompactSources} evaluates the likelihood that there is a large amount of emitting dust in sources that are very compact (and therefore do not intercept the light from distributed old populations like RGB stars and do not contribute to $\tau_{_{1.1}}$). Section \ref{SEC:Discussion} discusses possible interpretations of the results, and Section \ref{SEC:Conclusions} summarises our conclusions. 

Appendix \ref{APP:convert} explains why we work in terms of optical depth (rather than more intuitive and conventional metrics like the associated column-density of hydrogen). Appendix \ref{APP:NIRextinction} summarises the method used by \citet{Dalcetal2015} to estimate $\tau_{_{1.1}}$, and Appendix \ref{APP:FIRextinction} summarises the method used by \citet{Draietal2014} to estimate $\tau_{_{300}}$. Appendix \ref{APP:DustModels} presents a collection of theoretical dust models, for comparison with the properties of dust derived empirically in this paper.

\section{The standard procedure for analysing dust continuum emission}\label{SEC:StanProc}

\subsection{Basis of the standard procedure}

The standard procedure for analysing multi-wavelength maps of dust emission starts by degrading all  maps to the coarsest angular resolution (here, that of the longest wavelength, i.e. $\sim 36''$ at $500\,\mu\rm{m}$), thereby throwing away a large fraction of the available information. Then, it assumes that the emission is optically thin, and that there is a single type of dust, and a single dust temperature, along the line of sight through each pixel, so that the monochromatic intensity is
\begin{eqnarray}\label{EQN:Istandard}
I_{_\lambda}\!&\!=\!&\!\tau_{_\lambda}\;B_{_\lambda}(T)\;=\;\tau_{_{\lambda_{_{\rm o}}}}\left(\!\frac{\lambda}{\lambda\subO}\!\right)^{\!-\beta}\;B_{_\lambda}(T).
\end{eqnarray}
Here $\tau_{_\lambda}$ is the optical depth at wavelength $\lambda$; $B_{_\lambda}\!(T)$ is the Planck Function; $T$ is the dust temperature (as distinct from the gas kinetic temperature, which does not concern us in this paper); and $\lambda\subO$ is an arbitrary reference wavelength. $\beta$ reflects, to first order, how the dust opacity varies with wavelength in the far-IR, and hence the type of dust.

For pixels with good signal in all six {\it Herschel} wavebands \citep{Pogletal2010,Grifetal2010}, there is in principle sufficient information to estimate $\;\tau_{_{\lambda_{_{\rm o}}}}\!$, $\beta$ and $T$. However, low $T$ can be mimicked by high $\beta$ and vice versa, so many analyses fix $\beta=2$ (the value predicted by most theoretical dust models; see Appendix \ref{APP:DustModels}) and only fit $\tau_{_{\lambda_{_{\rm o}}}}$ and $T$. Given $\tau_{_{\lambda_{_{\rm o}}}}$, one can also estimate the surface density of dust, $\Sigma_{_{\rm DUST}}$, and the column density of hydrogen in all chemical forms, $N_{_{\rm H}}$. However, as noted in Appendix \ref{APP:convert}, these estimates introduce uncertain assumptions, and we do not need $\Sigma_{_{\rm DUST}}$ or $N_{_{\rm H}}$ here.

\subsection{Limitations of the standard procedure}

The main limitation of the standard procedure is that on most lines of sight there is a range of dust temperatures, basically because there is a wide range of radiation fields heating the dust; the more intense the ambient radiation field, the hotter the dust. And on many lines of sight there is a range of dust types, firstly because dust grains initially condense out under a range of different circumstances, and secondly because dust grains evolve according to the environment in which they find themselves; the denser and colder the environment, the more grains tend to grow, due to mantle accretion and/or coagulation. Therefore it is an oversimplification to assume that there is a single dust type, and a single dust temperature, along each line of sight.

The representative dust temperatures, ${\hat T}$, derived by the standard procedure are flux-weighted means. Since there is in reality a range of $T$, the contribution from warmer than average dust is overestimated, and the contribution from cooler than average dust is underestimated. The two errors do not in general cancel out.

Similarly, the representative emissivity indices, ${\hat\beta}$, derived by the standard procedure are also flux-weighted means. When there is in reality a range of $\beta$, the amount of cool dust with lower than average $\beta$ will be underestimated, and the amount of warm dust with lower than average $\beta$ will be overestimated. At the same time, the amount of cool dust with higher than average $\beta$ will be overestimated, and the amount of warm dust with higher than average $\beta$ will be underestimated.

Problems with the standard procedure become particularly severe when there are very small dust grains exposed to strong radiation fields. The very small dust grains are transiently heated. At any instant, most of the emission comes from a small subset of the grains that are briefly at extremely high temperatures and cooling rapidly.

\section{The {\sc PPMAP} procedure for analysing dust emission}\label{SEC:PPMAP}

\subsection{Basis of the {\sc ppmap} procedure}

As with the standard procedure, \pp also assumes that the dust emission is optically thin, and this can be checked retrospectively (see Section \ref{SEC:Correlations}). However, \pp does not assume a single uniform type of dust (uniform $\beta$), nor a single uniform dust temperature ($T$), along the line of sight through a pixel. \pp also delivers pixels which are $\sim 20$ times smaller in area than those delivered by the standard procedure.

\pp assumes that, on the line of sight through a pixel, the emitting dust has a continuous range of types (i.e. emissivity indices, $\beta$) and a continuous range of temperatures ($T$), and that these subscribe to a bivariate probability distribution, $P(\beta,T)$, so that the contribution to the total optical depth through the pixel at $\lambda\subO\!=\!300\,\mu{\rm m}$, $\tau_{_{300}}$, from dust with emissivity index in the interval $(\beta,\beta\!+\!d\beta)$ and temperature in the interval $(T,T\!+\!dT)$ is
\begin{eqnarray}
d^2\!\tau_{_{300}}&=&\tau_{_{300}}\;\frac{\partial^2\!P}{\partial\beta\,\partial T}\;d\beta\,dT\,.
\end{eqnarray}
By extension of Eqn. (\ref{EQN:Istandard}), the corresponding contribution to the monochromatic intensity in the pixel is
\begin{eqnarray}\nonumber
d^2\!I_{_\lambda}\!&\!=\!&\!d^2\!\tau_{_{300}}\,\left(\!\frac{\lambda}{\rm 300\,\mu m}\!\right)^{\!-\beta}B_{_\lambda}\!(T)\\
&\!=\!&\!\tau_{_{300}}\,\left(\!\frac{\lambda}{\rm 300\,\mu m}\!\right)^{\!-\beta}\,B_{_\lambda}\!(T)\;\frac{\partial^2\!P}{\partial\beta\,\partial T}\;d\beta\,dT\,,
\end{eqnarray}
and so the total monochromatic intensity in the pixel is
\begin{eqnarray}\nonumber
I_{_\lambda}\!&\!=\!&\!\int\limits_{\rm all\,\beta}\;\int\limits_{{\rm all}\,T}\tau_{_{300}}\,\left(\!\frac{\lambda}{\rm 300\,\mu m}\!\right)^{\!-\beta}\,B_{_\lambda}\!(T)\;\frac{\partial^2\!P}{\partial\beta\,\partial T}\;d\beta\,dT\,.\\\label{EQN:Ippmap.1}
\end{eqnarray}

\pp replaces the continuous ranges of $\beta$ and $T$ with a two-dimensional grid of discrete values, each representing a small but finite interval.  Specifically, for the analysis of M31, we define four linearly equal $\beta$-intervals between $1.25$ and $3.25$; hence the discrete values are $\beta_{_1}\!=\!1.5$,  $\beta_{_2}\!=\!2.0$,  $\beta_{_3}\!=\!2.5$,  $\beta_{_4}\!=\!3.0$, and each represents an interval $[\beta_{_k}\!-\!0.25,\beta_{_k}\!+\!0.25]$. Similarly, we define twelve logarithmically equal $T$-intervals between $9.3\,{\rm K}$ and $53.8\,{\rm K}$; hence the discrete values are $T_{_1}\!=\!10.0\,{\rm K}$, $T_{_2}\!=\!11.6\,{\rm K}$, $T_{_3}\!=\!13.4\,{\rm K}$, $T_{_4}\!=\!15.5\,{\rm K}$, $T_{_5}\!=\!18.0\,{\rm K}$, $T_{_6}\!=\!20.8\,{\rm K}$, $T_{_7}\!=\!24.1\,{\rm K}$, $T_{_8}\!=\!27.8\,{\rm K}$, $T_{_9}\!=\!32.2\,{\rm K}$, $T_{_{10}}\!=\!37.3\,{\rm K}$, $T_{_{11}}\!=\!43.2\,{\rm K}$, $T_{_{12}}\!=\!50.0\,{\rm K}$, and each represents an interval $[0.93T_{_\ell},1.08T_{_\ell}]$. The double integral in Eqn. (\ref{EQN:Ippmap.1}) can then be approximated by a double sum,
\begin{eqnarray}\nonumber
I_{_\lambda}&\simeq&\sum\limits_{k=1}^{k=4}\;\sum\limits_{\ell=1}^{\ell=12}\,\left\{\!\Delta^{\!2}\tau_{_{300:k\ell}}\left(\!\frac{\lambda}{\rm 300\,\mu m}\!\right)^{\!-\beta_{_k}}B_{_\lambda}(T_{_\ell})\!\right\}\!,\\
\end{eqnarray}
where $\Delta^{\!2}\tau_{_{300:k\ell}}$ is the contribution to $\tau_{_{300}}$ from dust with emissivity index in $\beta$-interval $k$ and temperature in $T$-interval $\ell$, i.e.
\begin{eqnarray}
\Delta^{\!2}\tau_{_{300:k\ell}}\!&\!=\!&\!\tau_{_{300}}\,\int\limits_{\beta=\beta_{_k}\!-0.25}^{\beta=\beta_{_k}\!+0.25}\;\,\int\limits_{T=0.93T_{_\ell}}^{T=1.08T_{_\ell}}\,\frac{\partial^2\!P}{\partial\beta\,\partial T}\;d\beta\,dT\,.
\end{eqnarray}
The raw data products of \pp are expectation values for $\Delta^{\!2}\tau_{_{300:k\ell}}$, and the corresponding uncertainties, $\Delta^{\!2}\sigma_{_{300:k\ell}}$, for the 48 combinations of $\beta_{_k}$ and $T_{_\ell}$ ($k\!=$1 to 4 times $\ell\!=$ 1 to 12), on the lines of sight through each of the pixels on M31 that has sufficient signal $(>5\sigma)$. We explain in Appendix \ref{APP:convert} why it is appropriate to formulate this problem in terms of optical depth, rather than the surface-density of dust, $\Sigma\subD$, or the associated column-density of gas, $N_{_{\rm H}}$.

\subsection{{\sc ppmap}'s underlying estimation procedure}\label{SEC:EstProc}

The \pp expectation values and uncertainties are derived using a Bayesian estimation procedure based on the concept of a {\it point process\/}, which is defined generically as the representation of a system as a collection of points in a suitably defined state space \citep{RichMars1991}. The system of interest here is the distribution and properties of dust in M31, which we represent with a  rectangular grid of cells, each occupied by an integer number of very small optical depth quanta, $\delta\tau_{_{300}}$. In the original formulation \citep{Marsetal2015}, each cell was described by just three parameters, namely its angular coordinates on the sky, $(x_{_i},y_{_j})$, and its dust temperature, $T_{_\ell}$, so that the ensemble of cells occupied a 3D state space  $(x,y,T)$. The procedure has since been enhanced to accommodate the emissivity index, $\beta$, so that the state space is now 4D, i.e. $(x,y,\beta,T)$, and the cells are distinguished by discrete values of $x_{_i},\;y_{_j},\;\beta_{_k}\;{\rm and}\;T_{_\ell}$ \citep{Marsetal2018}. The optical depth, $\;\Delta\!^2\tau_{_{300:k\ell}}$, assigned to a given cell is equal to the product of $\delta\tau_{_{300}}$ and the occupation number for that cell, $\Gamma_{_{\!ijk\ell}}$, i.e. the number of optical depth quanta, $\delta\tau_{_{300}}$, that have been allocated to that cell. The set of occupation numbers for all the cells is denoted by the state vector ${\mathbf\Gamma}$. For our analysis of M31 the number of pixels exceeds $10^6$, and on the lines of sight through each pixel there are $48$ combinations of $\beta$ and $T$, so the state vector has $\,\sim 5\times 10^7$ components.

The Bayesian estimation procedure is based on a measurement model of the form
\begin{equation}
{\mathbf d} = {\mathbf A}{\mathbf\Gamma} + {\mathbf\mu}\,.
\label{eq1}
\end{equation}
Here ${\mathbf d}$ is the measurement vector whose $m^{\rm th}$ component represents the pixel value at location $(X_m,Y_m)$ in the observed map at wavelength $\lambda_m$. $\;{\mathbf\mu}$ is the measurement noise, assumed to be a spatially and spectrally uncorrelated Gaussian random process with variance $\sigma_{\!\mu}^2$. $\;{\mathbf A}$ is the system response matrix whose $mn^{\rm th}$ element expresses the response of the $m^{\rm th}$ measurement to the optical depth, $\Delta^2\tau_{_{300:n}}$, in the $n^{\rm th}$ cell in the state space -- where the $n^{\rm th}$ cell corresponds to spatial location $(x_n,y_n)$, dust emissivity index $\beta_n$ and dust temperature $T_n$. $\;{\mathbf A}$ is given by
\begin{eqnarray}\nonumber
A_{_{mn}}\!&\!=\!&\!H_{_{\!\lambda_{_m}}}\!\!(X_m-x_n, Y_m-y_n)\,B\!_{_{\lambda_{_m}}}\!(T_{_n})\\\label{EQ:MeasurementModel}
&&\hspace{1.1cm}\times\;\Delta^2\tau_{_{300:n}}\;\left(\!\frac{\lambda_{_m}}{300\,\mu{\rm m}}\!\right)^{\beta_{_n}}\;\Delta\Omega_{_m}\,.
\label{eq0}
\end{eqnarray}
Here $H_{_\lambda}(x,y)$ is the convolution of the beam profile at wavelength $\lambda$ with the profile of an individual object, and $\Delta\Omega_m$ is the solid angle subtended by the $m^{\rm th}$ pixel.

\pp applies an iterative routine to obtain the set of expectation values for the cell occupation numbers, i.e. the components of the state vector ${\mathbf \Gamma}$. These are then scaled by $\delta\tau_{_{300}}$ to yield the differential optical depths, $\;\Delta\!^2\tau_{_{300:n}}$, and their corresponding uncertainties, $\;\Delta\!^2\sigma_{_{300:n}}$ \citep{Marsetal2015}. Note that for notational brevity we have condensed the grid of possible positions on the sky, $(x_{_i},y_{_j})$, possible emissivity indices, $\beta_{_k}$, and possible dust temperatures, $T_{_\ell}$, into a single index, $n$, representing a particular cell in the 4D state space. However, for the purpose of transforming the 4D image hypercube into projections (corresponding, for example, to images of mean $\beta$ or mean $T$), it is necessary to break out the index $n$ into $i,\;j,\;k\;{\rm and}\;\ell$ again, so that for a given spatial location, $(x_{_i},y_{_j})$, the optical depth in $\beta$-interval $k$ and $T$-interval $\ell$ is denoted $\Delta^2\tau_{_{300:k\ell}}$.  The iterative routine starts with all the occupation numbers set equal, and the noise level set -- {\it arbitrarily} -- so high that formally this is only a marginally unacceptable fit to the data. Hence the adjustments to the occupation numbers needed to improve the fit are sufficiently small to be in the linear regime. The linear adjustments are implemented, the noise level is reduced very slightly, and the process is repeated until the noise reaches the observed level. 

The observational noise at each wavelength is estimated by finding the standard deviation of sky background values in areas largely free of M31 emission. Iterations then proceed until the global value of reduced $\chi^2$ is just below 1, indicating that the model fitting errors are similar to the measurement noise.

The iterative routine is performed on small overlapping patches of the image field, and these patches are then stitched together so that all pixels on the final image incorporate the constraints that derive from their being coupled to neighbouring pixels by the point-spread function. Typically $\sim 2\times 10^4$ iterations are required for the patches on M31. Mathematical details of the iteration routine are given in \citet{Marsetal2015}.

\subsection{Advantages of {\sc ppmap}}

\begin{table}\begin{center}
\caption{Values of the global reduced $\chi^2$ and the reduced $\chi^2$s for the individual {\it Herschel} wavebands, along with the numbers of pixels, ${\cal N}_{_{\rm PIXEL}}$, that went into each value.}
\begin{tabular}{rclrcl}\hline
$\lambda/\mu{\rm m}$ & $\chi^2$ & ${\cal N}_{_{\rm PIXEL}}$ & $\hspace{1.0cm}\lambda/\mu{\rm m}$ & $\chi^2$ & ${\cal N}_{_{\rm PIXEL}}$ \\
70 & 0.9 & 4445743 & 350 & 1.3 & 437387 \\
100 & 0.8 & 4446276 & 500 & 1.3 & 200635 \\
160 & 0.7 & 2317468 & & & \\
250 & 1.1 & 822346 & {\sc global} & 0.86 & 12269855 \\\hline
\end{tabular}\label{TAB:chisquared}
\end{center}\end{table}

\pp achieves better resolution than the standard procedure because the measurement model (Eqn. \ref{EQ:MeasurementModel}) allows all the data to be used at their native resolution. For this work, we have used the resolution of the {\it Herschel} {\sc pacs} $70\,\mu{\rm m}$ map $(8'')$ to define the pixel size $(4'')$. Finer spatial resolution can in principle be invoked, but the uncertainties increase very rapidly if the spatial resolution is reduced below this value. The range of $\beta$-values considered, i.e. $(1.25,3.25)$, reflects the fact that most derived values of ${\bar\beta}$ fall in the range $(1.7,2.8)$. Similarly, the range of $T$-values considered, i.e. $(9.3\,{\rm K},53.8\,{\rm K})$, is dictated by the fact that most derived values of ${\bar T}$ fall in the range $(12\,{\rm K},18\,{\rm K})$, but with some much higher values in specific locations.

\pp could be run with additional, more closely spaced, discrete $\beta$ and/or $T$ values, but this would not actually increase the accuracy, and it would increase the required computing time. The choice of 4 discrete $\beta$ values and 12 discrete $T$ values is a compromise dictated by the amount of information in the input data, and the need to cover the inferred ranges of $\beta$ and $T$ (see preceding paragraph).

A further advantage of \pp is that it distinguishes dust of different types, and at different temperatures. This means that it gives more accurate values for the total optical depth than the standard procedure. In particular, \pp does not underestimate the amount of colder than average dust, or overestimate the amount of warmer than average dust, because it does not give all the dust on the line of sight a single representative temperature.

In addition to generating maps of the expectation value for the optical depth, $\;\Delta\!^2\tau_{_{300:k\ell}}$, and of the corresponding uncertainty, $\;\Delta\!^2\sigma_{_{300:k\ell}}$, at each combination of $\beta_{_k}$ and  $T_{_\ell}$, \pp produces synthetic {\it Herschel} maps internally and uses them to calculate the reduced $\chi^2$s for the individual {\it Herschel} wavebands, and also a global reduced $\chi^2$. The values obtained for M31 are given in Table \ref{TAB:chisquared}, along with the number of pixels (i.e. the number of independent data points) used to  obtain them.  

Finally, \pp is in principle able to handle the emission from small, transiently heated dust grains, provided that (a) the peak temperatures reached by transiently heated grains are not above the highest $T$-interval, and (b) the effective instantaneous emissivity index of a transiently cooling grain does not lie outside the available $\beta$-intervals. 

\begin{figure*}
\vspace{-1.5cm}
\hspace{-1.5cm}\includegraphics[width=0.865\textwidth]{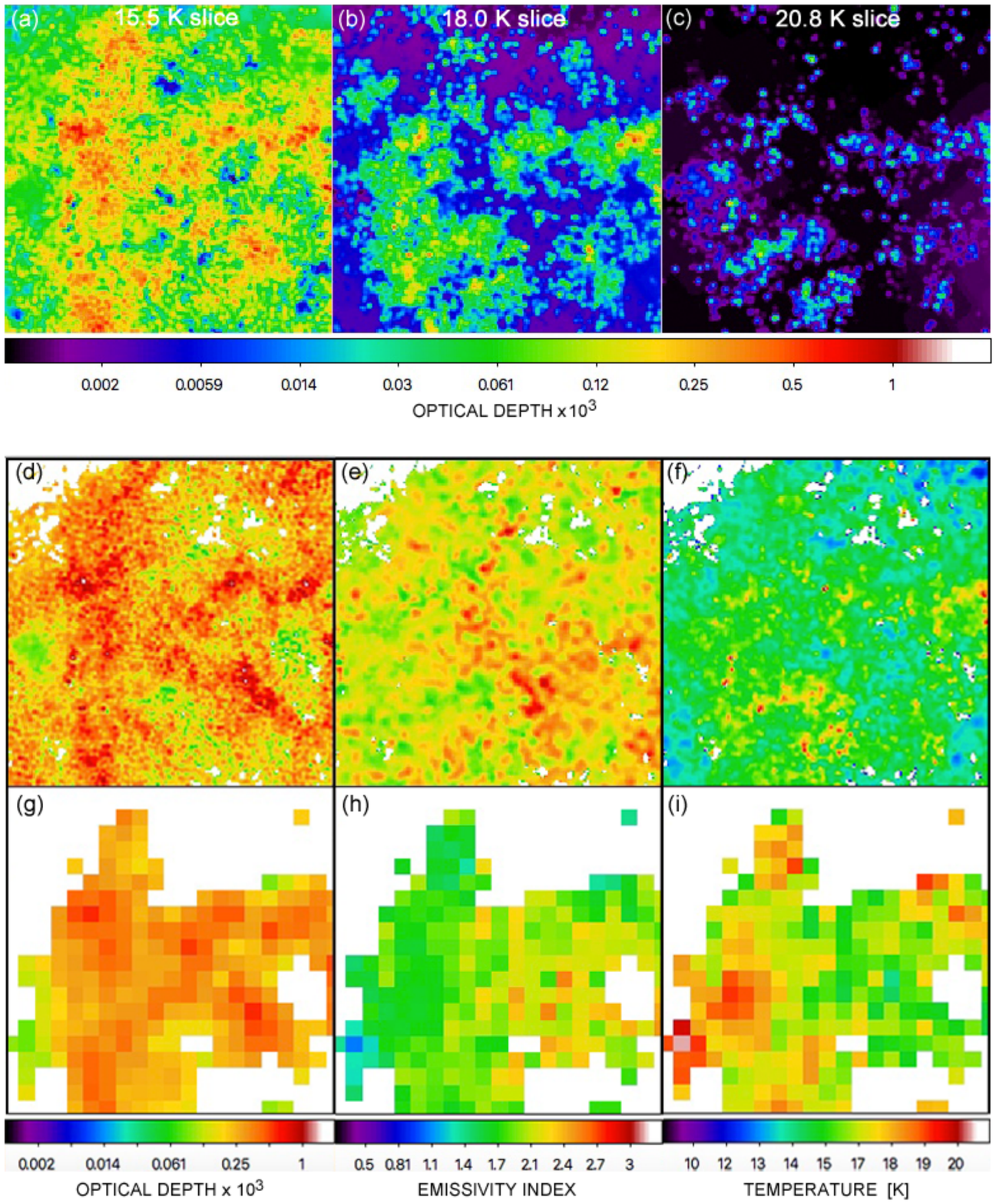}
\caption{Maps of the ZoomZone, a square $2.7\,{\rm kpc}\,\times\, 2.7\,{\rm kpc}$ region at the north-east extremity of the $11\,{\rm kpc}$ ring. The axes of the ZoomZone are aligned with equatorial coordinates: north is up, east to the left. Its centre is at ${\rm RA}=11.3499\,{\rm hr},\;{\rm Dec}=41.9050\deg$ (J2000). The ZoomZone is also marked with a black square on Fig. \ref{FIG:WholeGalaxy}(a). The first row shows temperature slices from three contiguous temperature intervals, (a) $\ell\!=\!4\;(14.4\,{\rm K}\,{\rm to}\,16.7\,{\rm K})$, (b) $\ell\!=\!5\;(16.7\,{\rm K}\,{\rm to}\,19.3\,{\rm K})$ and (c) $\ell\!=\!6\;(19.3\,{\rm K}\,{\rm to}\,22.4\,{\rm K})\;$ (i.e. images of  $\Delta\tau_{_{300:\ell}}$, as defined by Eqn. \ref{EQN:DeltaTauT}). The second row shows (d) the total optical depth, $\tau_{_{300}}$ (Eqn. \ref{EQN:Tau}); (e) the mean emissivity index, ${\bar\beta}$ (Eqn. \ref{EQN:BetaBar}); and (f) the mean dust temperature, $\bar{T}$ (Eqn. \ref{EQN:TBar}). The third row shows the corresponding images obtained with the standard procedure \citep{Smitetal2012}. Each image on the third row should be compared with the one immediately above it; further details are given in the text.}
\label{FIG:ZoomZone}
\end{figure*}

\subsection{Limitations of the current version of {\sc ppmap}}

The limitations of the current version of \pp are that (i) it delivers expectation values; (ii) it delivers no information about the distribution along the line of sight of the different types of dust or different dust temperatures; (iii) $\beta$-values may not be sufficient to discriminate between all types of dust; and (iv) it assumes that for all types of dust, $\beta$ is independent of $T$. The last two limitations can easily be relaxed, but this will only be sensible when better, i.e. more constraining, observations become available.

Because \pp delivers expectation values, the possibility exists that there is more than one significant peak in the a-posteriori probability distribution. This possibility seems unlikely, given the well-behaved nature of the functions involved in the response matrix (i.e. the Point Spread Function, Planck Function and far-IR emissivity law, see Eqn. \ref{EQ:MeasurementModel}), but it cannot be discounted. It is therefore reassuring that, as we discuss in Section \ref{SEC:Correlations}, the magnitude of the total optical depth, the mean emissivity index, the mean dust temperature, and their variations with galacto-centric radius all agree quite well with those obtained for M31 by \citet{Draietal2014} using a completely different procedure. 

\pp is not able to constrain where the dust of different types, and/or at different temperatures, lies along the line of sight, either in absolute terms (i.e. distances), or in relative terms (whether one type or temperature is behind, or in front of, another). This might be possible for a relatively unconfused line of sight, and given a simple model for the underlying distribution of dust, but the results would then be model dependent.

If there is more than one type of dust characterised by the same $\beta$, \pp can not, {\it  in its present form}, distinguish them; their contributions to the total optical depth are lumped together. However, given more sophisticated prescriptions for the wavelength dependence of the far-IR emissivities of different types of dust (i.e. more sophisticated than the single parameter $\beta$), it would be straightforward to adjust \pp to estimate the contributions from these different types.

Finally, in its present form, \pp assumes that for all dust types the emissivity, and hence $\beta$, is independent of the temperature, $T$. Again, it would be straightforward to adjust \pp so that this assumption could be relaxed.

\begin{figure*}
\hspace{-1.7cm}
\includegraphics[width=0.77\textwidth,angle=90]{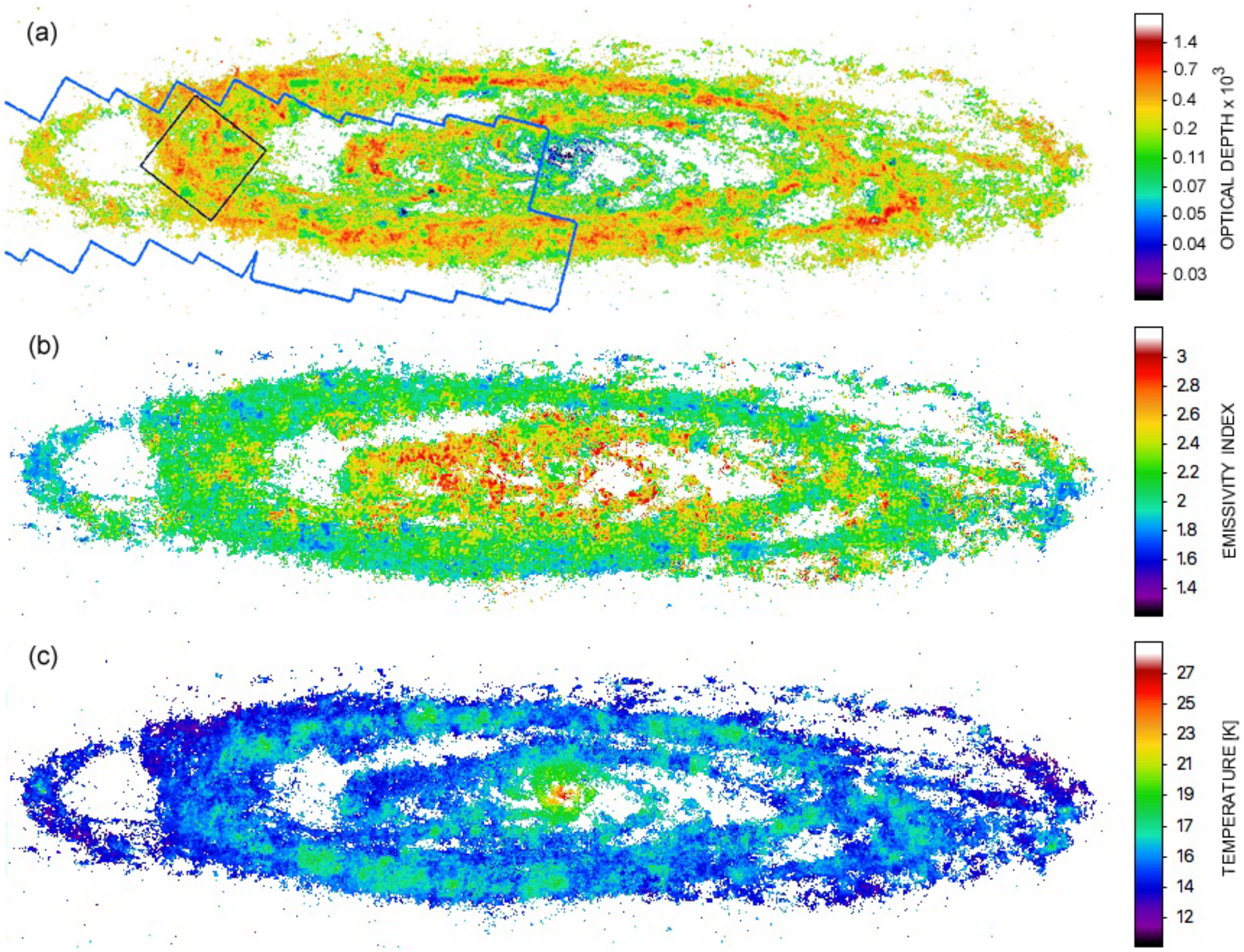}
\caption{\pp images of the whole of M31: (a) total far-IR optical depth at $300\,\mu{\rm m}$, $\tau_{_{300}}$, (b) mean emissivity index, ${\bar\beta}$, and (c) mean dust temperature, $\bar{T}$. On Panel (a), the black square delineates the region illustrated on Fig. \ref{FIG:ZoomZone}, and the blue outline delineates the sector analysed by \citet{Dalcetal2015}. These images have been rotated through $37.7^{\rm o}$ relative to the equatorial coordinate system.}
\label{FIG:WholeGalaxy}
\end{figure*}

\begin{figure*}
\hspace{-0.1cm}\vspace{-0.8cm}
\includegraphics[width=1.0\textwidth,angle=0]{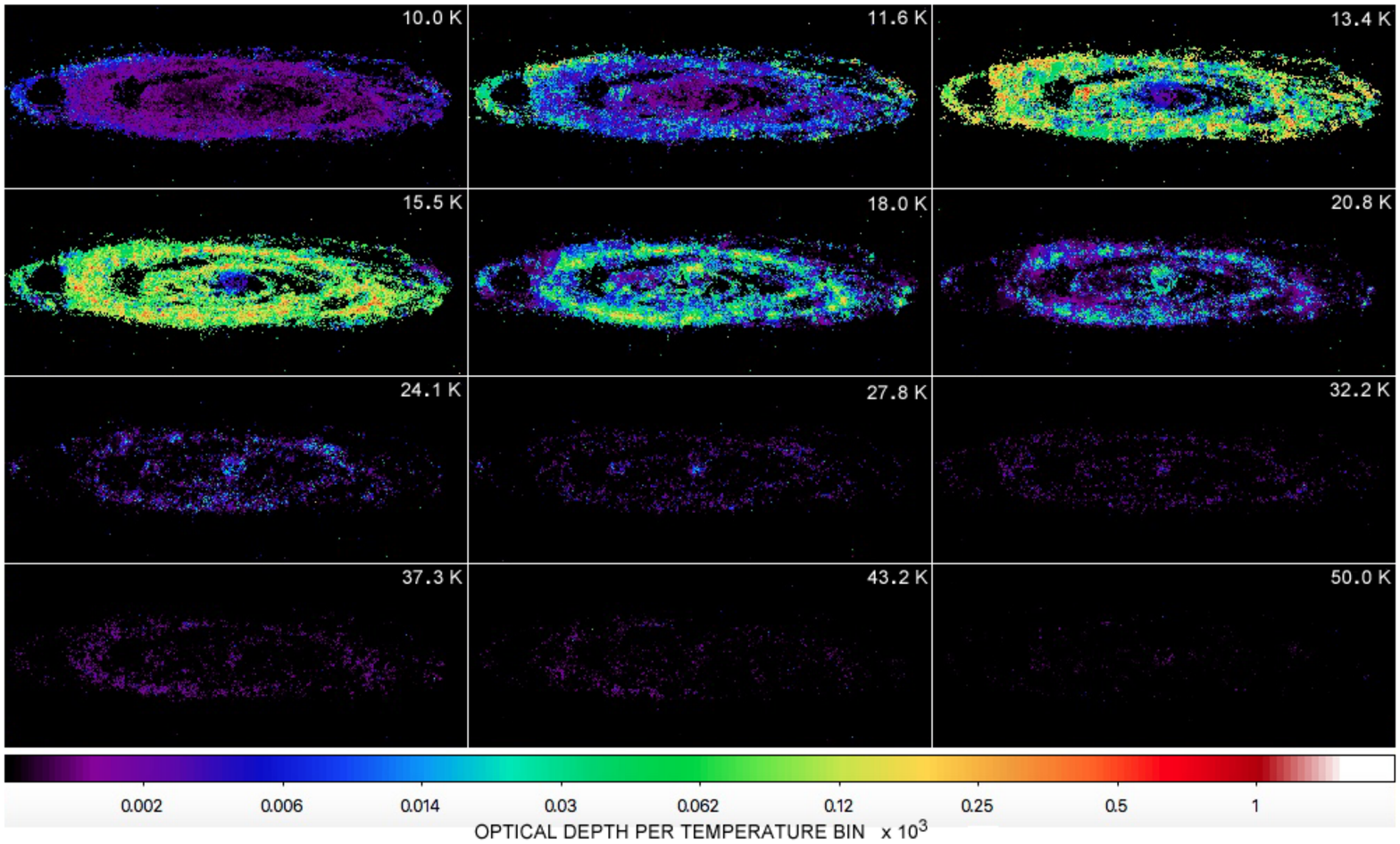}
\caption{Images of $\Delta\tau_{_{300:\ell}}$ (Eqn. \ref{EQN:DeltaTauT}), i.e. the contribution to the optical depth of dust at $300\,\mu{\rm m}$ from dust at the twelve discrete temperatures, $T_{_\ell}$, used by \ppp. On each panel, $T_{_\ell}$ is marked in the top right corner.}
\label{FIG:TempSlices}
\end{figure*}

\begin{figure*}
\hspace{-0.1cm}\vspace{-2.1cm}
\includegraphics[width=1.0\textwidth,angle=0]{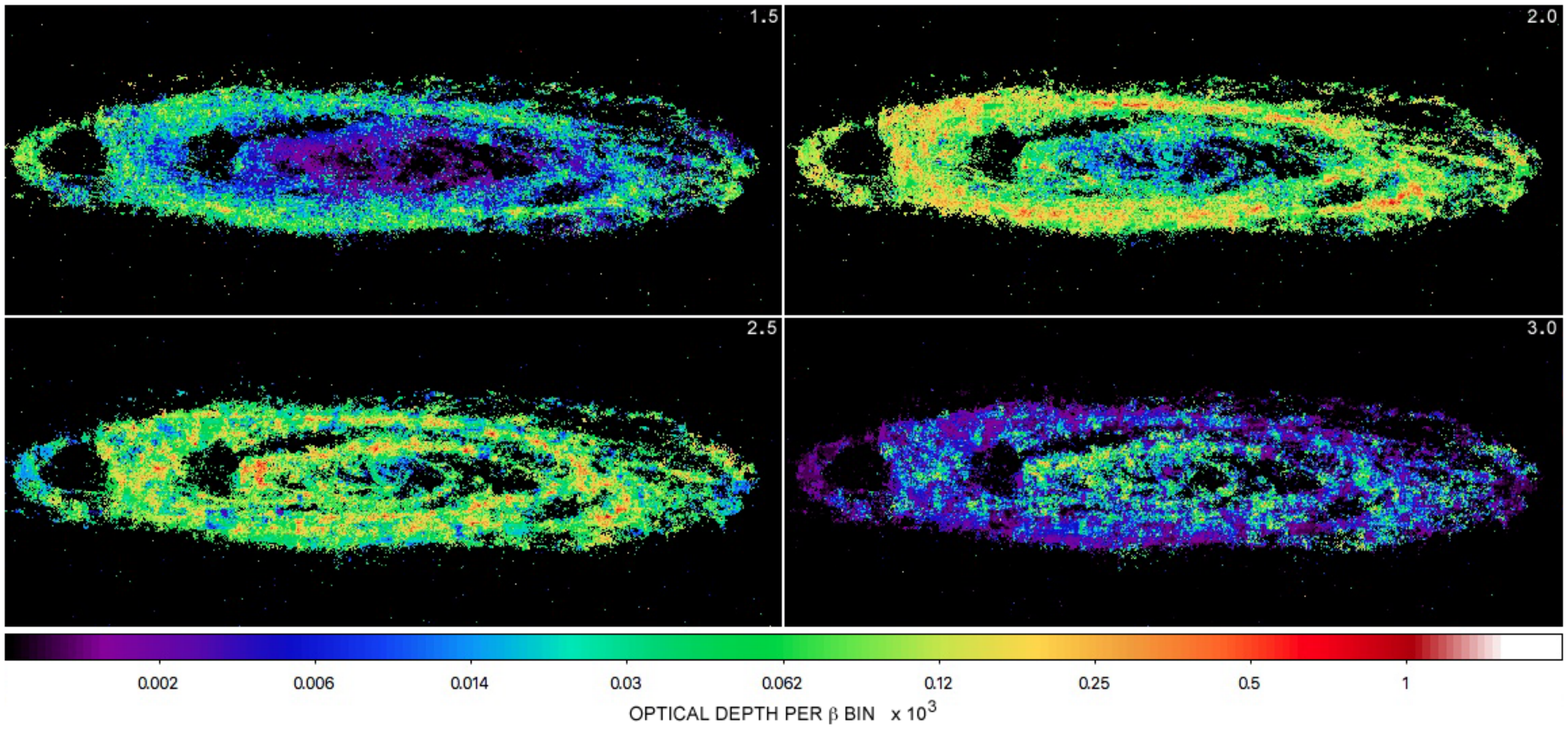}
\caption{Images of $\Delta\tau_{_{300:k}}$ (Eqn. \ref{EQN:DeltaTauBeta}), i.e. the contribution to the optical depth of dust at $300\,\mu{\rm m}$ from dust at the four discrete emissivity indices, $\beta_{_k}$, used by \ppp. On each panel, $\beta_{_k}$ is marked in the top right corner.}
\label{FIG:BetaSlices}
\end{figure*}

\section{Results} \label{SEC:Results}

To illustrate some of the \pp data products, we start by zooming in on a $2.7\,{\rm kpc}\,\times\, 2.7\,{\rm kpc}$ region at the north-east extremity of the $11\,{\rm kpc}$ ring. The location of this region (hereafter the {\it ZoomZone}) is marked with a square on the image of the whole of M31 on Fig. 2(a).

Given the values of $\Delta^{\!2}\tau_{_{300:k\ell}}$ for each pixel, we can compute a temperature slice for an individual $T$-interval, $\ell$, by summing over all the $\beta$-intervals, $k$,
\begin{eqnarray}\label{EQN:DeltaTauT}
\Delta\tau_{_{300:\ell}}&=&\sum\limits_{k=1}^{k=4}\left\{\Delta^{\!2}\tau_{_{300:k\ell}}\right\}.
\end{eqnarray}

The top row of Fig. \ref{FIG:ZoomZone} shows $T$-slices for the {\it ZoomZone} in three contiguous temperature intervals, $\ell\!=\!4\;(14.4\,{\rm K}\,{\rm to}\,16.7\,{\rm K})$, $\ell\!=\!5\;(16.7\,{\rm K}\,{\rm to}\,19.3\,{\rm K})$ and $\ell\!=\!6\;(19.3\,{\rm K}\,{\rm to}\,22.4\,{\rm K})$. These slices should be interpreted like velocity channel maps, where the velocity interval is replaced with a dust temperature interval, and the intensity (integrated over a velocity interval) is replaced with the optical depth (integrated over a temperature interval). The temperature slices therefore reveal how much dust (of all types) there is in the different $T$-intervals, and where it is located.

Similarly, emissivity index slices for individual $\beta$-intervals, $k$, can be computed by summing over all the $T$-intervals, $\ell$,
\begin{eqnarray}\label{EQN:DeltaTauBeta}
\Delta\tau_{_{300:k}}&=&\sum\limits_{\ell=1}^{\ell=12}\left\{\Delta^{\!2}\tau_{_{300:k\ell}}\right\}.
\end{eqnarray}
Emissivity index slices reveal how much dust (at all temperatures) there is in the different $\beta$-intervals. Hence they reveal where dust of different types is located.

The total optical depth is obtained by summing over both temperature (i.e. $\ell$) and emissivity index (i.e. $k$),
\begin{eqnarray}\label{EQN:Tau}
\tau_{_{300}}&=&\sum\limits_{\ell =1}^{\ell =12}\;\sum\limits_{k =1}^{k =4}\left\{\Delta^{\!2}\tau_{_{300:k\ell}}\right\}.
\end{eqnarray}
The optical depth weighted mean emissivity index and mean temperature are then given by
\begin{eqnarray}\label{EQN:BetaBar}
{\bar\beta}&=&\frac{1}{\tau_{_{300}}}\;\;\sum\limits_{\ell =1}^{\ell =12}\;\sum\limits_{k =1}^{k =4}\left\{\;\beta_{_k}\;\Delta^{\!2}\tau_{_{300:k\ell}}\,\right\},\\
\label{EQN:TBar}
{\bar T}&=&\frac{1}{\tau_{_{300}}}\;\;\sum\limits_{\ell =1}^{\ell =12}\;\sum\limits_{k =1}^{k =4}\left\{\,T_{_\ell}\;\Delta^{\!2}\tau_{_{300:k\ell}}\,\right\}.
\end{eqnarray}
From the internal error model, and from simulations using synthetic data, we find that the {\it absolute} uncertainty on ${\bar\beta}$ is $\sim\!0.1$, and the {\it fractional} uncertainty on ${\bar T}$ is $\sim\! 0.03$.\footnote{We note that, if, for example, all the dust on the line of sight through a particular pixel, say $(i,j)\!=\!(42,57)$, had $\beta\!=\!(\beta_{_2}+\beta_{_3})/2=2.25$ and $T\!=\!(T_{_6}T_{_7})^{1/2}=22.4\,{\rm K}$, \pp would allocate comparable amounts of optical depth to the cells $(i,j,k,\ell)=(42,57,2,6),\;(42,57,2,7),\;(42,57,3,6)\;{\rm and}\;(42,57,3,7)$, and hence return ${\bar\beta}\sim 2.25$ and ${\bar T}\sim 22.4\,{\rm K}$.}

The middle row of Fig. \ref{FIG:ZoomZone} shows, reading from left to right, (d) the total optical depth, $\tau_{_{300}}$ (Eqn. \ref{EQN:Tau}); (e) the optical-depth weighted mean emissivity index, ${\bar\beta}$ (Eqn. \ref{EQN:BetaBar}); and (f) the optical-depth weighted mean dust temperature, ${\bar T}$ (Eqn. \ref{EQN:TBar}), in the {\it ZoomZone}.

The third row of Fig. \ref{FIG:ZoomZone} shows the corresponding results obtained using the standard analysis procedure \citep{Smitetal2012} on the {\it ZoomZone}: reading from left to right, (g) a single notional optical-depth, ${\hat\tau}_{_{300}}$; (h) a single notional emissivity index, ${\hat\beta}$; and (i) a single notional temperature, ${\hat T}$.\footnote{Throughout the paper, we use ${\bar\beta}$ and ${\bar T}$ to denote optical-depth weighted averages along the line of sight, based on {\sc ppmap} data products. We use ${\hat\beta}$, ${\hat T}$ and ${\hat\tau}$ to denote the flux-weighted averages derived by Smith et al. (2012) using the standard procedure. And we use ${\tilde\beta}$, ${\tilde T}$ and ${\tilde\tau}$ to denote the quantities derived by Draine et al. (2014) using their irradiation algorithm.} In all nine panels of Fig. \ref{FIG:ZoomZone}, only pixels with $5\sigma$ significance are populated.

The pixels obtained with \pp are approximately twenty times smaller in area than those obtained with the standard procedure. Moreover, the properties evaluated within the \pp pixels are better defined, because we have the distribution of dust as a function of both $\beta$, and $T$, in 48 $(\beta_{_k},T_{_\ell})$ combinations. By applying \pp and the standard procedure to synthetic data, we have shown that \pp delivers more accurate, and sometimes significantly different, optical-depths \citep{Marsetal2015}. In particular, \pp registers both colder than average dust (which, with the standard procedure, gets lost in the glare from warmer dust) {\it and} hotter than average dust (which, with the standard procedure, can lead to the mass of dust being overestimated).

Fig. \ref{FIG:WholeGalaxy} shows images of (a) $\tau_{_{300}}$, (b) ${\bar\beta}$, and (c) ${\bar T}$, obtained with \pp for the whole of M31 (the same quantities as Panels \ref{FIG:ZoomZone}d, \ref{FIG:ZoomZone}e and \ref{FIG:ZoomZone}f, which only cover the region within the black square on Panel \ref{FIG:WholeGalaxy}a). \citet{Smitetal2012} have analysed {\it Herschel} maps of M31 using the standard procedure (see Section \ref{SEC:StanProc}), which delivers a resolution of $\sim\!140\,{\rm pc}\;(\sim 36'')$ . \citet{Draietal2014} have analysed {\it Herschel} maps of M31 using a sophisticated irradiation algorithm that also exploits {\sc Spitzer} data to constrain emission from transiently heated grains and the role of very strong local radiation fields (see Appendix \ref{APP:FIRextinction}), and they achieve a resolution of $\sim\!90\,{\rm pc}\;(\sim 23'')$. With $4''$ pixels \pp delivers a resolution of $\sim\!31\,{\rm pc}\;(\sim 8'')$, sufficient to start to resolve Giant Molecular Clouds, and to evaluate correlations between dust properties and environment.

Fig. \ref{FIG:TempSlices} shows the twelve individual temperature slices generated by \ppp, i.e. the contributions, $\Delta\tau_{_{300:\ell}}$ (Eqn. \ref{EQN:DeltaTauT}), to the total optical depth, $\tau_{_{300}}$ (Eqn. \ref{EQN:Tau}), from the twelve discrete  dust temperatures, $T_{_\ell}$. Each map should be interpreted as the contribution to $\tau_{_{300}}$ from dust in a small interval about $T_{_\ell}$; for example the map at $T_{_2}=11.6\,{\rm K}$ actually represents dust in the interval $10.8\,{\rm K}\la T\la 12.5\,{\rm K}$. These maps show that most of the dust is in the range between $\sim 12\,{\rm K}$ and $\sim 20\,{\rm K}$, with the warmest dust concentrated in the centre and in star formation regions in the $11\,{\rm kpc}$ ring.

Fig. \ref{FIG:BetaSlices} shows the four individual emissivity-index slices generated by \ppp, i.e. the contributions, $\Delta\tau_{_{300:k}}$ (Eqn. \ref{EQN:DeltaTauBeta}), to the total optical depth, $\tau_{_{300}}$ (Eqn. \ref{EQN:Tau}), from the four discrete emissivity indices, $\beta_{_k}$. Each map should be interpreted as the contribution to $\tau_{_{300}}$ from dust in a small interval about $\beta_{_k}$; for example the map at $\beta_{_2}=2.0\,{\rm K}$ actually represents dust in the interval $1.75\,{\rm K}\la \beta\la 2.25\,{\rm K}$. These maps show that most of the dust in M31 has $1.75\la\beta\la 2.75$; dust with $\beta\la 1.75$ is concentrated towards the outer parts of M31 ($r\ga 11\,{\rm kpc}$), and most of the dust with $\beta\ga 2.75$ is concentrated towards the  centre ($r\la 5\,{\rm kpc}$).

\begin{figure*}
\includegraphics[width=0.8\textwidth]{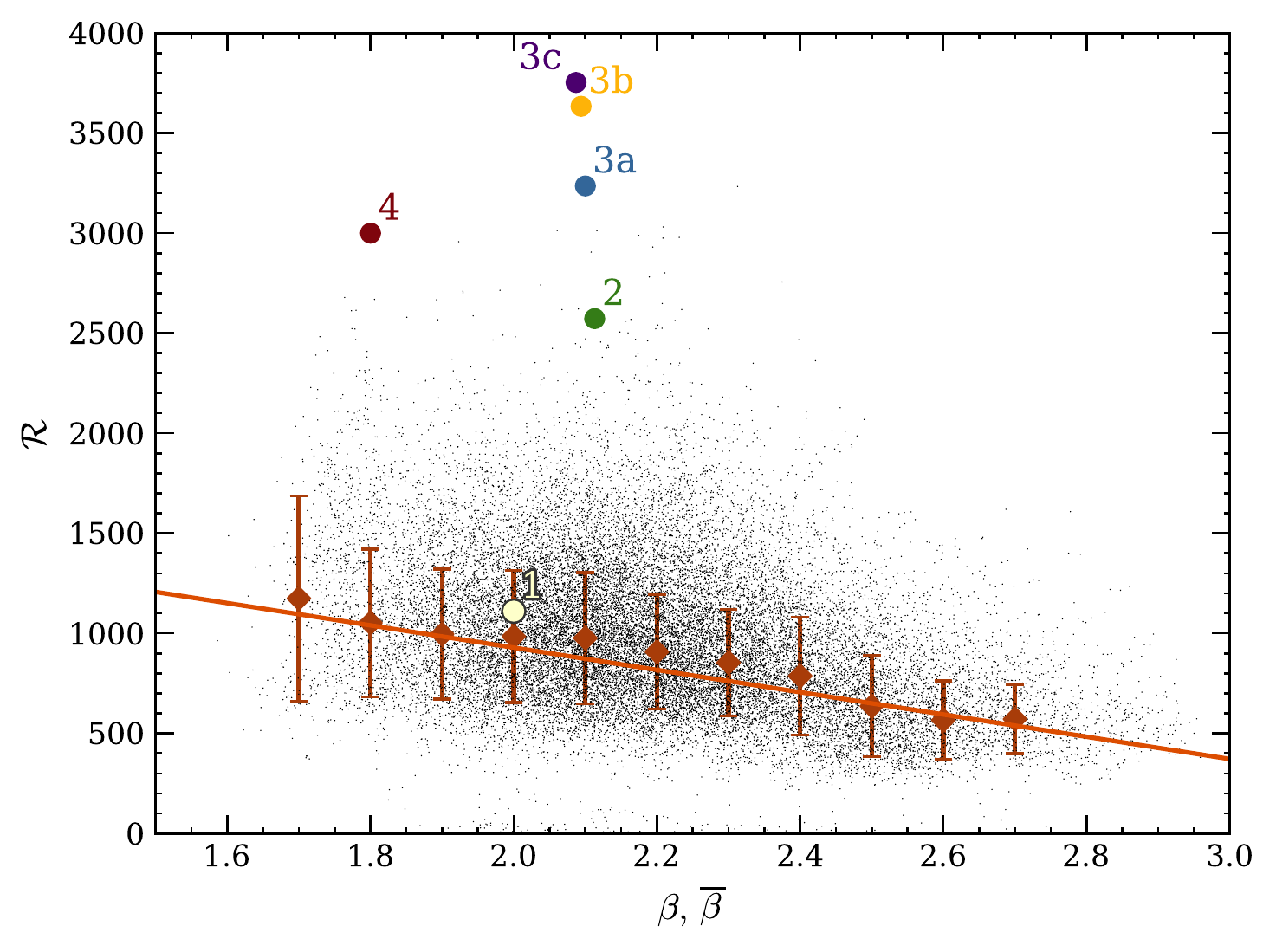}
\caption{Plot of ${\cal R}^{\mbox{\tiny obs.}}_\tau\!=\!\tau_{_{1.1}}/\tau_{_{300}}$ (Eqn. \ref{EQN:OpticalDepthRatio}) against ${\bar\beta}$ (Eqn. \ref{EQN:BetaBar}) for the 28726 \pp pixels that have robust ($>\!\!5\sigma$) detections; each pixel is represented by a small black dot. The red diamonds and error bars show the means and standard deviations in finite bins, ${\bar\beta}\pm 0.05$. The red line is the linear regression fit (Eqn. \ref{EQN:RbetaFit}) to the individual pixel points. For comparison, the filled circles represent values of ${\cal R}^{\mbox{\tiny model}}_\kappa\!=\!\kappa_{_{1.1}}/\kappa_{_{300}}$ (Eqn. \ref{EQN:OpacityRatio}) and $\beta$ from different theoretical dust models, with the associated number, or number and letter, giving the source reference, as listed in Table \ref{TAB:DustModels1}; further details are given in Appendix \ref{APP:DustModels}.}
\label{FIG:calR-beta}
\end{figure*}

\begin{table*}
\begin{center}
\caption{Tabulated models from the literature. Columns 1 and 2 give the values of $\beta$ and ${\cal R}^{\mbox{\tiny model}}_\kappa$. Columns 3 and 4 give a brief indication of the model ingredients and the source reference. Column 5 gives the ID used to identify these models on Figs. \ref{FIG:calR-beta} and \ref{FIG:calR-beta_X}.}
\begin{tabular}{lllll}\hline
$\beta\hspace{1.0cm}$ & ${\cal R}^{\mbox{\tiny model}}_\kappa\hspace{0.5cm}$ & {\sc Model Ingredients}$\hspace{1.7cm}$ & {\sc Source}$\hspace{2.2cm}$ & {\sc ID} \\
2.00 & 1111 & mainly observation &  \citet{MathisJS1990} & 1 \\
2.11 & 2573 & a-C, graphite, a-Sil & \citet{LiDraine2001} & 2 \\
2.10 & 3236 & a-C, graphite, a-Sil; $R_{_{\rm V}}\!=\!3.1$ & \citet{DraineBT2003} & 3a \\
2.09 & 3634 & a-C, graphite, a-Sil; $R_{_{\rm V}}\!=\!4.0$ & \citet{DraineBT2003} & 3b \\
2.09 & 3753 & a-C, graphite, a-Sil; $R_{_{\rm V}}\!=\!5.5$ & \citet{DraineBT2003} & 3c \\
1.80 & 3000 & a-C, a-C(:H), a-Sil$_{\rm Fe}$ & \citet{Joneetal2013} & 4 \\\hline
\end{tabular}\label{TAB:DustModels1}
\end{center}
\end{table*}

\section{Correlations}\label{SEC:Correlations}

The $31\,{\rm pc}$ resolution of the image of $\tau_{_{300}}$ obtained with \pp (our Fig. \ref{FIG:WholeGalaxy}a) is close to the $25\,{\rm pc}$ resolution of the image of the near-infrared extinction optical depth at $1.1\,\mu{\rm m}$, $\;\tau_{_{1.1}}$, obtained from the reddening statistics of Red Giant Branch (RGB) stars in the north-east sector of M31 by Dalcanton et al. (2015; their Fig. 21). There is also close morphological correspondence between the two images. We can therefore evaluate the ratio of optical depths at these two wavelengths,
\begin{eqnarray}\label{EQN:OpticalDepthRatio}
{\cal R}^{\mbox{\tiny obs.}}_\tau&=&\frac{\tau_{_{1.1}}}{\tau_{_{300}}}\,,
\end{eqnarray}
as a function of position, over the region treated by \citet{Dalcetal2015}. This region, hereafter the {\it Overlap Region}, is outlined in blue on Fig. \ref{FIG:WholeGalaxy}(a). Strictly speaking, we are comparing the {\it extinction} optical depth at $1.1\,\mu{\rm m}$ with the {\it absorption/emission} optical depth at $300\,\mu{\rm m}$, but since the albedo of dust at $300\,\mu{\rm m}$ is presumed to be negligible, we can treat both as extinction optical depths. 

Fig. \ref{FIG:calR-beta} shows a plot of ${\cal R}^{\mbox{\tiny obs.}}_\tau$ (Eqn. \ref{EQN:OpticalDepthRatio}) against ${\bar\beta}$ (Eqn. \ref{EQN:BetaBar}). All 28726 pixels in the Overlap Region that have reliable optical depths at both wavelengths are represented by small black points. The red line on Fig. \ref{FIG:calR-beta} is a linear fit to these points, 
\begin{eqnarray}\label{EQN:RbetaFit}
{\cal R}^{\mbox{\tiny obs.}}_\tau&\simeq& 2042\,(\pm 24)\;-\;557\,(\pm 10)\,{\bar\beta}\,,
\end{eqnarray}
and the red diamonds with error bars represent the means and standard deviations in contiguous bins ${\bar\beta}\pm 0.05$ for ${\bar\beta}\!=\!1.7,\,1.8,\,.\,.\,.\;2.6,\,2.7$.

For comparison, the filled circles on Fig. \ref{FIG:calR-beta} show values of
\begin{eqnarray}\label{EQN:OpacityRatio}
{\cal R}^{\mbox{\tiny model}}_\kappa&=&\frac{\kappa_{_{1.1}}}{\kappa_{_{300}}}
\end{eqnarray}
for several commonly used theoretical dust models. Here, $\kappa_{_{1.1}}$ is the near-IR extinction opacity at $1.1\,\mu{\rm m}$; $\kappa_{_{300}}$ is the far-IR extinction opacity at $300\,\mu{\rm m}$; and the models are listed in Table \ref{TAB:DustModels1}, along with the IDs used to distinguish the filled circles on Fig. \ref{FIG:calR-beta}.

As already noted by \citet{Dalcetal2015} -- and with the exception of the \citet{MathisJS1990} model -- the theoretical values of ${\cal R}^{\mbox{\tiny model}}_\kappa$ exceed the observed values of ${\cal R}^{\mbox{\tiny obs.}}_\tau$ by at least a factor of order $2.5$. This discrepancy  \citep[which is probably related to the `dust energy balance problem', e.g.][]{Saftetal2015} was also noted by {\sc Planck} \citep{PlanckXIV2014}.

There are (at least) three possible explanations for the discrepancy. {\it Explanation A:} the analyses used to evaluate $\tau_{_{300}}$ --- here, \ppp; and in \citet{Draietal2014}, the irradiation algorithm outlined in Appendix \ref{APP:FIRextinction} --- may be giving the wrong answer; we argue below that, since the \ppp-based analysis presented here and the irradiation algorithm used by \citet{Draietal2014} arrive at similar answers, by completely different routes, this is unlikely. {\it Explanation B:} it may be that a significant fraction of the dust emitting at $300\,\mu{\rm m}$ is in configurations which are so compact that they very seldom intercept the lines of sight to background RGB stars on the far side of M31; in Section \ref{SEC:CompactSources} we present two analytic arguments which indicate that this is unlikely. {\it Explanation C:} it may be that new dust models are needed; if this is the case then the correlations that we derive below may provide useful constraints on the constitution of interstellar dust, and how it responds to different environments.

\begin{figure*}
\hspace{0.0cm}
\includegraphics[trim=0.0mm 14.0mm 0.0mm 10.0mm,clip=true,width=0.95\textwidth]{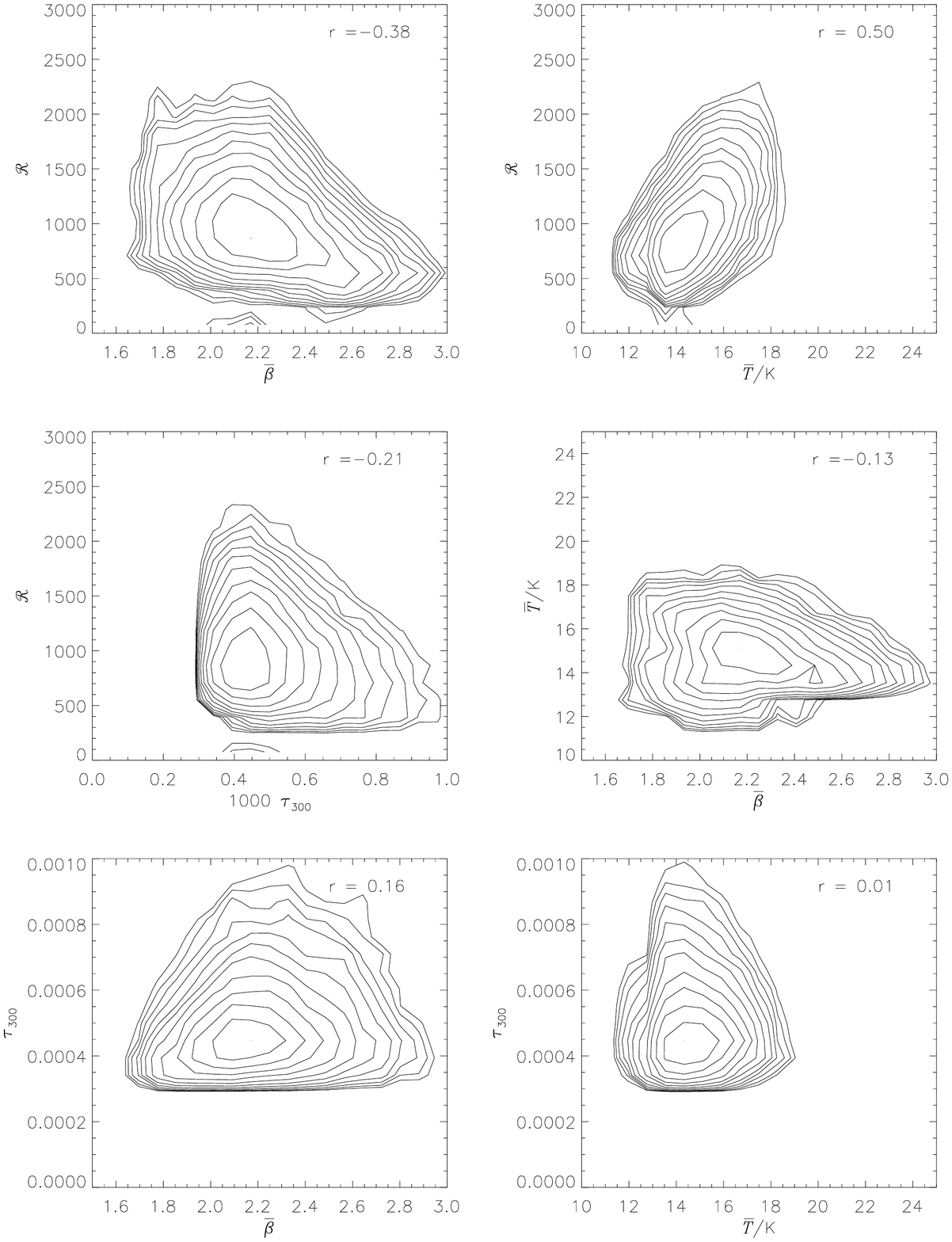}
\caption{Correlations between the values of $\tau_{_{300}}$, ${\bar\beta}$, $\bar{T}$ and ${\cal R}^{\mbox{\tiny obs.}}$ in all pixels where there is a robust $(>\!5\sigma)$ signal. The Pearson correlation coefficients are marked in the top righthand corner of each panel. Contours go down from the peak, $N_{_{\rm PEAK}}$, by successive factors of $2^{1/2}$, and the outermost contour is at $\sim 0.022N_{_{\rm PEAK}}$.}
\label{FIG:Correlations}
\end{figure*}

Fig. \ref{FIG:Correlations} presents the correlations between $\tau_{_{300}}$, ${\bar\beta}$, ${\bar T}$ and ${\cal R}^{\mbox{\tiny obs.}}_\tau$. The sharp lower limit on $\tau_{_{300}}$ derives from the fact that lower values do not get past our $5\sigma$ cut. ${\cal R}^{\mbox{\tiny obs.}}$ is correlated with ${\bar T}$, but anti-correlated with ${\bar\beta}$ and $\tau_{_{300}}$. ${\bar T}$ is anti-correlated with ${\bar\beta}$, but only very mildly. $\tau_{_{300}}$ is weakly correlated with ${\bar\beta}$, but un-correlated with ${\bar T}$. 

\begin{figure*}
\includegraphics[trim=10.0mm 0.0mm 6.0mm 0.0mm,clip=true,width=0.65\textwidth,angle=90]{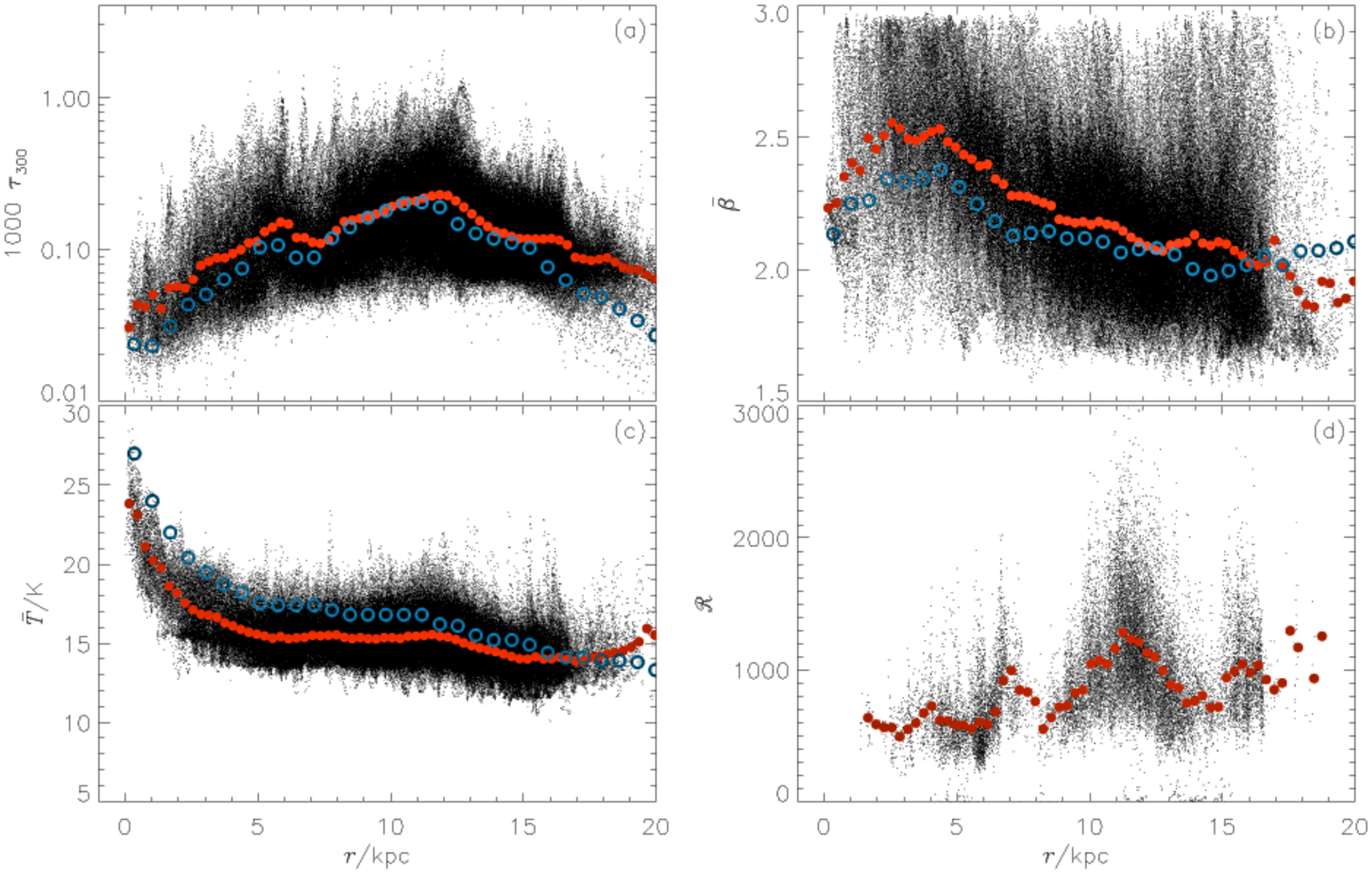}
\caption{Radial profiles of (a) the total optical depth at $300\,\mu{\rm m}$, $\tau_{_{300}}(r)$; (b) the mean emissivity index, ${\bar\beta}(r)$; (c) the mean dust temperature, $\bar{T}(r)$; and (d) the ratio of optical depths at $1.1\,\mu{\rm m}$ and $300\,\mu{\rm m}$, ${\cal R}^{\mbox{\tiny obs.}}_\tau(r)$, where $\;r$ is galactocentric radius. The small black dots correspond to individual pixels, and the filled red circles are azimuthal averages in annuli with width $\Delta r=300\,{\rm pc}$; for comparison, the open blue circles show the results obtained by \citet{Draietal2014} with $\Delta r=677\,{\rm pc}$. There are fewer points on Panel c because ${\cal R}^{\mbox{\tiny obs.}}_\tau$ can only be evaluated where there are estimates of $\tau_{_{1.1}}$ from \citet{Dalcetal2015}, i.e. in the north-east sector.}
\label{FIG:Profiles}
\end{figure*}

Fig. \ref{FIG:Profiles} presents the variations of $\tau_{_{300}}(r)$, ${\bar\beta}(r)$, ${\bar T}(r)$ and ${\cal R}^{\mbox{\tiny obs.}}_\tau(r)$ with galacto-centric radius, $r$. The small black dots represent individual pixels, and the filled red circles show azimuthal averages in annuli of width $\Delta r=300\,{\rm pc}$. For comparison, the open blue circles show the azimuthal averages obtained by \citet{Draietal2014} in annuli with $\Delta r=677\,{\rm pc}$. We should be mindful (a) that \citet{Draietal2014} used a completely different procedure from us to obtain their results, with lower spatial resolution; (b) that our radial profiles only extend to $r\!\sim\!20\,{\rm kpc}$, whereas those in \citet{Draietal2014} extend to $r\!\sim\!25\,{\rm kpc}$; and (c) that the \pp results are essentially model independent.

Our Fig. \ref{FIG:Profiles}(a) should be compared with Fig. 3(b) from \citet{Draietal2014}. To make this comparison, we have converted their deprojected dust surface density, $\Sigma_{_{\rm M,dust}}\cos(i)$, into our un-deprojected dust optical depth, $\,\tau_{_{300}}=\Sigma_{_{\rm M,dust}}\kappa_{_{300}}$. Here $i\!=\!77.7^{^{\rm o}}$ is the inclination angle between M31's midplane and the plane of the sky, hence $\cos(i)\!=\!0.21$, and $\kappa_{_{300}}\!=\!2.7\,{\rm cm^2\,g^{-1}}$ is the mass opacity coefficient at $300\,\mu{\rm m}$. Consequently $\tau_{_{300}}=2.7\times 10^{-9}\,[\Sigma_{_{\rm M,dust}}\cos(i)/({\rm M}_{_\odot}{\rm kpc}^{-2})]$. In general, and in particular where the results are most robust (between $\sim 2\,{\rm kpc}$ and $\sim 15\,{\rm kpc}$), there is reasonable correspondence between our results and theirs, both as regards absolute values of $\tau_{_{300}}$, and as regards radial variations,  for example the minimum between $6\,{\rm kpc}$ and $8\,{\rm kpc}$ and the maximum near $11\,{\rm kpc}$.

Our Fig. \ref{FIG:Profiles}(b) should be compared with Fig. 13 from \citet{Draietal2014}. This comparison is somewhat compromised by the fact that \citet{Draietal2014} define $\beta$ in a post-processing step, between $250\,\mu{\rm m}$ and $500\,\mu{\rm m}$. In contrast, we define $\beta$ as an intrinsic parameter of the \pp analysis, across the entire wavelength range, i.e. between $70\,\mu{\rm m}$ and $500\,\mu{\rm m}$. Our ${\bar\beta}$ has a slightly larger dynamical range, $1.9\la{\bar\beta}\la 2.5$, as compared with their $2.0\la{\bar\beta}\la 2.4$, but the overall trends are similar. One should expect a somewhat increased dynamic range, given that \pp has finer resolution. 

Our Fig. \ref{FIG:Profiles}(c) should be compared with Fig. 9(b) from \citet{Draietal2014}. Our values of ${\bar T}$ are systematically lower than those obtained by \citet{Draietal2014}, but the radial variation obtained by the two analyses is similar.

Appendix \ref{APP:FIRextinction} gives a brief description of the analysis procedure used by \citet{Draietal2014} to estimate the dust parameters of M31, and in particular to estimate $\Sigma\subD$. This procedure is very different from \ppp. In particular, \pp invokes no model assumptions, neither concerning the radiation field, nor concerning the dust (beyond the assumption that the variation of the long-wavelength opacity with wavelength can be approximated with an emissivity index, $\beta$). The agreement in the radial profiles, in particular regarding $\tau_{_{300}}$, is an indication that both procedures are physically sound, and that the results they obtain are credible. We are therefore inclined to dismiss {\it Explanation A}. 

Our Fig. \ref{FIG:Profiles}(d) does not have an equivalent in \citet{Draietal2014}, because the near-IR $\,1.1\,\mu{\rm m}$ optical depths from \citet{Dalcetal2015} were not available to \citet{Draietal2014} and so ${\cal R}^{\mbox{\tiny obs.}}_\tau$ could not be evaluated. The main inference from Fig. \ref{FIG:Profiles}(d) is that the higher values of ${\cal R}^{\mbox{\tiny obs.}}_\tau$ are concentrated in the dense star-forming rings. However, we should also note that the reason there are fewer pixel-points from the lines of sight between the rings is because optical depths there are lower, and therefore many pixels fail to meet the $5\sigma$ threshold applied to both the \pp parameters and those derived by \citet{Dalcetal2015}.

From Figs. 6 and 7 we see that $\tau_{_{300}}\stackrel{<}{\sim} 0.001$, and hence, even with $\beta =3.0$, $\;\tau_{_{70}}\stackrel{<}{\sim} 0.08\,.$ Therefore the assumption that the emission is optically thin appears to be valid.

\section{Very compact emission sources}\label{SEC:CompactSources}

{\it Explanation B} requires that -- unless we adopt  the \citet{MathisJS1990} dust model -- a large fraction of the dust emitting in the far-IR is in sources which are so compact that they are unlikely to intercept the lines of sight to RGB stars on the far side of M31. Specifically, the requirement is that a fraction
\begin{eqnarray}
{\cal F}&=&\frac{{\cal R}^{\mbox{\tiny model}}_\kappa-{\cal R}^{\mbox{\tiny obs.}}_\tau}{{\cal R}^{\mbox{\tiny model}}_\kappa}\;\;=\;\;1-\left(\frac{{\cal R}^{\mbox{\tiny model}}_\kappa}{{\cal R}^{\mbox{\tiny obs.}}_\tau}\right)^{\!-1}
\end{eqnarray}
of the emitting dust be located in these very compact sources. Substituting ${\cal R}^{\mbox{\tiny model}}_\kappa\ga 2.5\,{\cal R}^{\mbox{\tiny obs.}}_\tau$, we obtain ${\cal F}\ga 0.6$. Below we present two analyses which suggest that this is unlikely, and therefore that {\it Explanation B} may not be tenable. The first analysis (Section \ref{SEC:NPDF}) is based on an evaluation of the consequences for the observed column-density PDF; and the  second analysis (Section \ref{SEC:SFR}) on an evaluation of the consequences for the rate of star formation.

\subsection{Consequences of very compact sources for the tail of the column-density PDF}\label{SEC:NPDF}

The near-IR extinction optical depths are obtained by \citet{Dalcetal2015} on the assumption that in each pixel there is a log-normal distribution of extinctions, and hence, by implication, a log-normal distribution of column-densities, $\Sigma$, characterised by a median, ${\tilde\Sigma}$, and a variance, $\sigma\simeq0.35\pm 0.10$. We hypothesise that, in addition to the log-normal distribution, there is, on most lines of sight, a power-law tail extending to much higher values of surface-density, and characterised by a parameter $\phi$ (measuring how far below its peak, the log-normal is intercepted by the power-law tail) and an exponent $-\alpha$. If we define $\eta=\Sigma/{\tilde\Sigma}$, the distribution of $\eta$ values can be approximated by
\begin{eqnarray}\label{EQN:dpdlambda}
\frac{dP}{d\eta}&=&\left\{\begin{array}{ll}
K_{_{\rm O}}\,,\hspace{2.0cm} & -\sigma <\eta <+\sigma\,; \\
K_{_{\rm O}}\,\phi\;{\rm e}^{-\alpha\eta}\,, & +\sigma\leq\eta <\infty\,.
\end{array}\right.
\end{eqnarray}
For mathematical convenience, the Gaussian shape of the log-normal has been approximated with a box-car; this is the first expression on the righthand side of Eqn. \ref{EQN:dpdlambda}. In the same spirit, the power-law tail, the second expression on the righthand side of Eqn. \ref{EQN:dpdlambda}, has been extended to infinity; strictly speaking, it should be limited to $\eta$ values for which the far-IR dust emission is optically thin, but these values are so large that setting the limit on $\eta$ to infinity makes no significant difference. 

We can now compute the ratio of the probabilities that a random line of sight intercepts the power-law tail (PT), or the log-normal (LN; {\it vice} box-car),
\begin{eqnarray}
\frac{P_{_{\rm PT}}}{P_{_{\rm LN}}}&\simeq&\frac{\phi}{2\,\sigma\,\alpha\;{\rm e}^{\alpha\sigma}}\,.
\end{eqnarray}
We can also compute the ratio of the corresponding masses,
\begin{eqnarray}\label{EQN:Mratio}
\frac{M_{_{\rm PT}}}{M_{_{\rm LN}}}&\simeq&\frac{\phi}{(\alpha-1)\,{\rm e}^{\alpha\sigma}\,\left(1-{\rm e}^{-2\sigma}\right)}\,.
\end{eqnarray}
In the interests of simplicity, we assume that the dust in the very compact sources of the PT has the same temperature as the more widely distributed dust of the LN; in this case Eqn. (\ref{EQN:Mratio}) also gives the ratio of the dust luminosities, $L_{_{\rm PT}}/L_{_{\rm LN}}$, and we require $M_{_{\rm PT}}/M_{_{\rm LN}}\ga 0.6$. In reality, the dust in the very compact sources of the PT is observed to be cooler than the more widely distributed dust of the LN \citep[e.g.][]{Marsetal2015}, so we should expect $M_{_{\rm PT}}/M_{_{\rm LN}}>L_{_{\rm PT}}/L_{_{\rm LN}}$. In this case, the lower limit on $M_{_{\rm PT}}/M_{_{\rm LN}}$ is even greater than $0.6$. This would make the conclusion that we reach below even stronger.

If we now set $\sigma\!=\!0.25$ \citep[a lower than average value according to][]{Dalcetal2015}, and require (a) that $P_{_{\rm PT}}/P_{_{\rm LN}}\la 0.1$ (i.e. fewer than $10\%$ of lines of sight to RGB stars go through the power-law tail, so they might have been missed), and (b) that $M_{_{\rm PT}}/M_{_{\rm LN}}\ga 0.6$ (i.e. at least $60\%$ of the dust emission is from the power-law tail), we must have $\phi\la 0.5$ and $\alpha\la 1.5$. In other words, we require a very shallow tail which intercepts the log-normal above the half-maximum point. If we increase $\sigma$ to $0.45$ \citep[a higher than average value according to][]{Dalcetal2015}, the lower limit on $\phi$ increases (the tail intercepts the log-normal even closer to its peak) and the upper limit on $\alpha$ decreases (the tail becomes even shallower still).  

Observed column-density PDFs from massive star-forming regions very occasionally do have power-law tails satisfying these conditions \citep{Schnetal2015b,Schnetal2015a}. However, many more lines of sight have  power-law tails with much smaller $\phi$ and much larger $\alpha$, and even more lines of sight have no discernible power-law tails at all. We conclude that there does not appear be a power-law tail to the distribution of column-densities in M31 that can deliver sufficient extra compact long-wavelength dust emission.

\begin{figure*}
\hspace{0.0cm}
\includegraphics[width=1.0\textwidth]{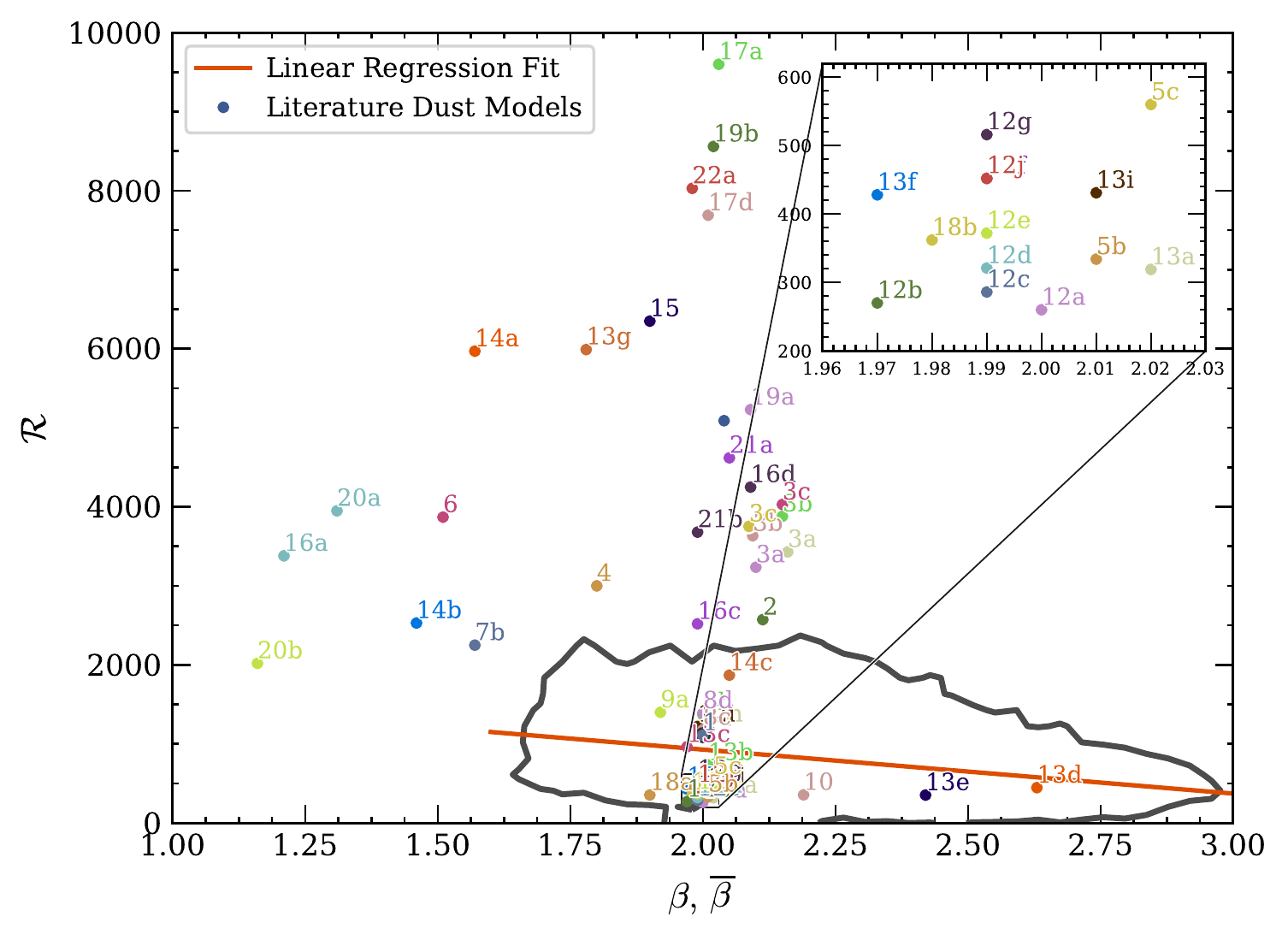}
\caption{The filled circles give values of $\beta$ and ${\cal R}^{\mbox{\tiny model}}_\kappa$ for the tabulated dust models from Table \ref{TAB:DustModels1} (IDs 1 to 4), and for the single-size models from Table \ref{TAB:DustModels2} (IDs 5a to 22b); further details of these models are given in Appendix \ref{APP:DustModels}. The black line encloses $90\%$ of the values of ${\bar\beta}$ and ${\cal R}^{\mbox{\tiny obs.}}_\tau$ for the 28726 \pp pixels on the {\it Herschel} image with robust values ($>5\sigma$), and the red line is the linear regression fit to these values (Eqn. \ref{EQN:RbetaFit}).}
\label{FIG:calR-beta_X}
\end{figure*}

\subsection{Consequences of very compact sources for the star formation rate}\label{SEC:SFR}

An alternative approach to estimating the contribution of compact sources to the long-wavelength dust emission is to consider a population of dense cores created by turbulence, as in the theory of turbulent star formation \citep{PadoNord2002}. In this theory, the distribution of core masses, $m$, can be approximated by
\begin{eqnarray}\label{EQN:dNdMcore}
\frac{d{\cal N}}{dm}&\simeq&K_{_{\rm O}}\,\left(\!\frac{m}{\rm{M}_{_\odot}}\!\right)^{\!-7/3}\!,\hspace{0.5cm}m_{_{\rm MIN}}\la m\la m_{_{\rm MAX}}\,.
\end{eqnarray}
Strictly speaking we should set $m_{_{\rm MAX}}\sim 100\,{\rm M}_{_\odot}$, since more massive cores are so extended that they could not fail to intercept the lines of sight from background RGB stars, but we will set $m_{_{\rm MAX}}$ to infinity, since this makes the analysis simpler and strengthens our final conclusion. The most critical parameter here is $m_{_{\rm MIN}}$.

In the turbulent theory of star formation, essentially all the high-mass cores spawn high-mass stars, but proceeding to lower masses, fewer and fewer cores get compressed enough to become gravitationally unstable and spawn low mass stars and brown dwarfs -- hence the turn-over in the Initial Mass Function. There should therefore be a large population of low-mass non-prestellar cores. From Eqn. (\ref{EQN:dNdMcore}), the total mass of the core population is 
\begin{eqnarray}
{\cal M}_{_{\rm TOT}}&\sim& 3\,K_{_{\rm O}}\,{\rm M}_{_\odot}^2\,\left(\!\frac{m_{_{\rm MIN}}}{{\rm M}_{_\odot}}\!\right)^{\!-1/3}.
\end{eqnarray}
If this is to exceed $\sim\!60\%$ of the gas mass in M31, i.e. ${\cal M}_{_{\rm TOT}}\ga\,4\times 10^9\,{\rm M}_{_\odot}$, we must have  
\begin{eqnarray}\label{EQN:K_MIN}
K_{_{\rm O}}&\ga&1.3\times 10^9\;{\rm M}_{_\odot}^{-1}\,\left(\!\frac{m_{_{\rm MIN}}}{{\rm M}_{_\odot}}\!\right)^{\!1/3}\,.
\end{eqnarray}

We can obtain a second constraint on $K_{_{\rm O}}$ by considering only those high-mass cores (say $m\ga 30\,{\rm M}_{_\odot}$) that form high-mass stars (say $m_\star\ga 8\,{\rm M}_{_\odot}$). The expectation is that virtually all these cores spawn high-mass stars, because they are almost always gravitationally unstable. In the Milky Way, the rate of high-mass star formation is $\;\la0.01\,\rm{yr}^{-1}$, and in M31 it is probably lower. Moreover, the time for a high-mass star to condense out of a high-mass core is $\la10^7\rm{yr}$. Therefore the number of high-mass cores in M31 should satisfy ${\cal N}_{_{>30{\rm M}_\odot}}\la10^5$. From Eqn. (\ref{EQN:dNdMcore}) the number of high-mass cores is 
\begin{eqnarray}
{\cal N}_{_{>30{\rm M}_\odot}}&\simeq& 0.75\,K_{_{\rm O}}\,{\rm M}_{_\odot}\left(\frac{30{\rm M}_{_\odot}}{{\rm M}_{_\odot}}\right)^{-4/3},
\end{eqnarray}
so ${\cal N}_{_{>30{\rm M}_\odot}}\la10^5$ requires
\begin{eqnarray}\label{EQN:K_MAX}
K_{_{\rm O}}&\la&1.3\times 10^5\;{\rm M}_{_\odot}^{-1}\,\left(\!\frac{30\,{\rm M}_{_\odot}}{{\rm M}_{_\odot}}\!\right)^{\!4/3}\,.
\end{eqnarray}
Combining Eqns. (\ref{EQN:K_MIN}) and (\ref{EQN:K_MAX}), we obtain
\begin{eqnarray}
m_{_{\rm MIN}}\!&\!\la\!&\!10^{-12}\,{\rm M}_{_\odot}\!\left(\!\frac{30\,{\rm M}_{_\odot}}{{\rm M}_{_\odot}}\!\right)^{\!4}\simeq\;8\times 10^{-7}\,{\rm M}_{_\odot},
\end{eqnarray}
which is of order a quarter the mass of the Earth. This would require $35\%$ of the mass of the interstellar medium to be in non-prestellar cores less massive than the Earth, and $90\%$ to be in non-prestellar cores less massive than Jupiter. We conclude that low-mass non-prestellar cores are unlikely to provide enough long-wavelength emission to explain the discrepancy between ${\cal R}^{\mbox{\tiny obs.}}_\tau$ and ${\cal R}^{\mbox{\tiny model}}_\kappa$.

\section{Discussion}\label{SEC:Discussion}

If {\it Explanations A} and {\it B} for the discrepancy between ${\cal R}^{\mbox{\tiny obs.}}_\tau$ and ${\cal R}^{\mbox{\tiny model}}_\kappa$ are hard to uphold (as argued in Sections \ref{SEC:Correlations} and \ref{SEC:CompactSources} respectively) we may need to consider {\it Explanation C} seriously. The inference is that some dust models may have to be abandoned, but also that new models may be required, and we suggest some constraints on such models.

In order to broaden the context within which dust models may need to be revised, Fig. \ref{FIG:calR-beta_X} shows both the tabulated dust models from Table \ref{TAB:DustModels1} that were already plotted on Fig. \ref{FIG:calR-beta}, and the single-size models from Table \ref{TAB:DustModels2}; the latter have been computed using Mie Theory with optical constants from the literature, and further details are given in Appendix \ref{APP:DustModels}. The red line on Fig. \ref{FIG:calR-beta_X} is the best fit to the anti-correlation between ${\cal R}^{\mbox{\tiny obs.}}_\tau$ and ${\bar\beta}$ (Eqn. \ref{EQN:RbetaFit}), and the black contour contains $90\%$ of the 28726 individual pixel-points plotted on Fig. \ref{FIG:calR-beta}. Almost all the models lie near or above the red line, and near or to the left of a second {\it undrawn} line that goes through $({\bar\beta},{\cal R}^{\mbox{\tiny obs.}}_\tau)\!\sim\!(2.0,1000)$ and is approximately orthogonal to the red line.

There is likely to be more than one type of dust in the interstellar medium of M31. Moreover, lines of sight through the disc of M31 will often intercept different phases of the interstellar medium, and the mix of dust types in these different phases is expected to vary. The derived values of ${\cal R}$ and ${\bar\beta}$ are therefore very unlikely to correspond to a single type of dust; they are optical depth weighted means of all the dust types along the line of sight. However, they must fall on the $({\cal R},\beta)$ plane inside the convex hull of the points representing the different constituent dust types, and close to those points that represent the dominant dust types. Figs. \ref{FIG:calR-beta} and \ref{FIG:calR-beta_X} then impose rather stringent constraints on the mix of dust models in M31.

The simplest way to explain the red line would be to invoke two types of dust one at the lefthand end, and one at the righthand end, with different proportions of these two types of dust on different lines of sight. Although this is certainly an over-simplification, it indicates where the search for relevant dust models might start. First, models are needed that deliver $(\beta,{\cal R}^{\mbox{\tiny model}}_\kappa)\!\sim\!(2.0,1000)$, like \citet{MathisJS1990}, or possibly even further up the red line on Fig. \ref{FIG:calR-beta_X}, i.e. even smaller $\beta$ and somewhat higher ${\cal R}^{\mbox{\tiny model}}_\kappa$. Second, models are needed that deliver $(\beta,{\cal R}^{\mbox{\tiny model}}_\kappa)\!\sim\!(2.5,500)$, or further down the red line on Fig. \ref{FIG:calR-beta_X}. From Fig. \ref{FIG:Profiles}(d), it appears that models delivering higher than average ${\cal R}^{\mbox{\tiny model}}_\kappa$ should be concentrated in the rings, and therefore presumably in denser than average gas or close to newly-formed luminous stars.

When comparing these results with those obtained previously for M31, and for other nearby galaxies, we should be mindful of the fact that \pp delivers unprecedented resolution on M31 ($15\,{\rm pc}$ pixels), and estimates the distribution of dust over a range of emissivity indices ($\beta$) and temperatures ($T$). Consequently \pp is likely to find more extreme values for these parameters, since previous analyses have necessarily been limited to averages over the line of sight and/or over larger areas.

In M31, {\sc Planck} \citep{Planck252015} obtains $\sim 1\,{\rm kpc}$ resolution, and finds a range $1.4\la{\hat\beta}\la 2.4$ (with mean 1.6), and a range $12\,{\rm K}\la{\hat T}\la 23\,{\rm K}$ (with mean $18\,{\rm K}$). \citet{Smitetal2012} obtain $\sim 140\,{\rm pc}$ resolution, and find ranges $1.2\la{\hat\beta}\la 2.8$ and $14\,{\rm K}\la{\hat T}\la 30\,{\rm K}$. \citet{Draietal2014} obtain $\sim 90\,{\rm pc}$ resolution, but average over annuli with width $\Delta r=677\,{\rm pc}$, and find ranges $1.9\la{\tilde\beta}\la 2.5$ and $12\,{\rm K}\la{\tilde T}\la 32\,{\rm K}$. With \pp we obtain $\sim 31\,{\rm pc}$ resolution, and find ranges $1.7\la{\bar\beta}\la 3.0$ (with mean 2.2), and $12\,{\rm K}\la{\bar T}\la 27\,{\rm K}$ (with mean $16\,{\rm K}$).

For the {\sc Kingfish} sample of nearby galaxies, \citet{Kirketal2014} find ranges $0.85\la{\hat\beta}\la 2.25$ and $16\,{\rm K}\la {\hat T}\la 30\,{\rm K}$ for the cool dust; they also include a warm dust component with a fixed temperature of $60\,{\rm K}$ in their models. For M33, \citet{Tabaetal2014} obtain $\sim 160\,{\rm pc}$ resolution, and obtain ranges $1.2\la{\hat\beta}\la 1.8$ and $18\,{\rm K}\la{\hat T}\la 23\,{\rm K}$ when they fit pixels with a single-component model, and $0.8\la{\hat\beta}\la 2.3$ and $16\,{\rm K}\la{\hat T}\la 60\,{\rm K}$ when they fit pixels with a double-component model. Unlike us, \citet{Tabaetal2014} find higher values of ${\hat\beta}$ in the star formation regions. In the Magellanic Clouds, \citet{Gordetal2014} obtain $\sim 12\,{\rm pc}$ resolution and find ranges $1.0\la{\hat\beta}\la 2.5$ and $15\,{\rm K}\la{\hat T}\la 30\,{\rm K}$. In the local Milky Way, {\sc Planck} \citep{Planck482016} finds ranges $1.3\la{\hat\beta}\la 1.9$ (with mean 1.6), and $17\,{\rm K}\la{\hat T}\la 22\,{\rm K}$ (with mean $19.4\,{\rm K}$).

All these results suggest the need for dust models with a wide range of $\beta$ values. Many seem to require models with $\beta >2.2$, and the \pp results suggest that these models may have ${\cal R}^{\mbox{\tiny model}}_\kappa\sim 500$.

\section{Conclusions}\label{SEC:Conclusions}

We have presented and analysed images of the dust in M31 obtained by applying \pp to {\it Herschel} far-IR data; and we have evaluated three possible explanations for the apparent discrepancy between the optical depth of dust required by the far-IR emission and the optical depth required to explain the reddening of RGB stars on the far side of M31. The main technical results and inferences are:
\begin{enumerate}
\item \pp delivers images with $\sim\!8''$ resolution, essentially corresponding to the shortest {\it Herschel} wavelength, $70\,\mu{\rm m}$.
\item This corresponds to $\sim\!31\,{\rm pc}$ at the distance of M31, which is on the order of the scale of a Giant Molecular Cloud.
\item \pp delivers separate images for the expectation value of the far-IR ($300\,\mu{\rm m}$) dust emission optical depth, $\tau_{_{300}}$, in different intervals of emissivity index ($\beta$) and different intervals of dust temperature ($T$). 
\item In principle, this allows \pp to calculate the total far-IR optical depth, $\tau_{_{300}}$, more accurately (than the standard procedure), because the amount of warmer than average dust is not overestimated by according it too low a temperature, and the amount of cooler than average dust is not underestimated by according it too high a temperature.
\item \pp also delivers separate images for the uncertainty in the dust optical depth in different $\beta$-intervals and different $T$-intervals.
\item From the \pp data products we can compute, in each $4''\!\times\! 4''$ pixel, the optical-depth weighted mean emissivity index, ${\bar\beta}$, and the optical-depth weighted mean dust temperature, ${\bar T}$.
\item Images of the near-IR ($1.1\,\mu{\rm m}$) dust extinction optical depth, $\tau_{_{1.1}}$, obtained by \citet{Dalcetal2015} from the reddening of RGB stars on the far side of M31's disc, have a similar resolution ($\sim\! 25\,{\rm pc}$) to our far-IR images ($\sim\!31\,{\rm pc}$).
\item Consequently we are able to compute ${\cal R}^{\mbox{\tiny obs.}}_\tau\equiv\tau_{_{1.1}}/\tau_{_{300}}$ on the scale of our \pp pixels.
\item The evaluation of ${\cal R}^{\mbox{\tiny obs.}}_\tau$ is almost entirely empirical. The derivation of $\tau_{_{1.1}}$ only assumes that the distribution of dust optical depths in M31 can be fit with a log-normal and that the scale-height of the dust in M31 is much less than that of the RGB stars. The derivation of $\tau_{_{300}}$ only assumes that the far-IR dust opacity can be fit with a power law (i.e. $\beta$), and that the far-IR emission is optically thin.
\end{enumerate}
The main science results and inferences are:
\begin{enumerate}
\item ${\cal R}^{\mbox{\tiny obs.}}_\tau$ derived in this way is significantly smaller than the values of ${\cal R}^{\mbox{\tiny model}}_\kappa\equiv\kappa_{_{1.1}}/\kappa_{_{300}}$ (where $\kappa_{_{L}}$ is the dust opacity at wavelength $L\,\mu{\rm m}$) for most commonly used theoretical dust models; the one exception is the model of \citet{MathisJS1990}. This is a variant on an already well established discrepancy between dust observations and dust theory (see Section \ref{SEC:Correlations}).
\item ${\cal R}^{\mbox{\tiny obs.}}_\tau$ is anti-correlated with ${\bar\beta}$, according to ${\cal R}^{\mbox{\tiny obs.}}_\tau\simeq 2042(\pm 24)-557(\pm 10){\bar\beta}$ (Eqn. \ref{EQN:RbetaFit}). This appears to be a new result that may help in identifying the shortcomings of existing dust models; even the \citet{MathisJS1990} model does not explain the high-$\beta$ end of this correlation (see Fig. \ref{FIG:calR-beta}).
\item One possible explanation for the discrepancy between ${\cal R}^{\mbox{\tiny obs.}}_\tau$ and ${\cal R}^{\mbox{\tiny model}}_\kappa$ is that the \pp results are inaccurate; this seems unlikely, given that they agree so closely with the results obtained by \citet{Draietal2014} using a completely different analysis procedure (see Section \ref{SEC:Correlations}). 
\item A second possible explanation for the discrepancy is that a significant fraction ($\ga 60\%$) of the dust emitting in the far-IR is located in such compact configurations that it is unlikely to intercept the lines of sight from RGB stars on the far side of M31; we present two lines of reasoning that suggest this is extremely unlikely (see Section \ref{SEC:CompactSources}). 
\item A third possible explanation is that new dust models are required.
\item These new models must explain the values of $({\bar\beta},{\cal R}^{\mbox{\tiny obs.}}_\tau)\!\sim\!(2.0,1000)$, which currently are only fit by the \citet{MathisJS1990} models.
\item They must also explain the values of $({\bar\beta},{\cal R}^{\mbox{\tiny obs.}}_\tau)\!\sim\!(2.5,500)$, which are not explained by any of the commonly used models.
\item If interstellar dust has low values of ${\cal R}^{\mbox{\tiny model}}_\kappa\la 1000$, the implication is that $\kappa_{_{300}}$ must be increased by $\ga 2.5$. In turn, this will reduce the dust masses of external galaxies, where these have been derived from their far-IR fluxes, which will relax somewhat the need for rapid dust formation in high-redshift galaxies \citep{Dunnetal2003,MorgEdmu2003}.
\end{enumerate}

\section*{Acknowledgements}

APW, KAM, MWLS, OL, MJG and SAE gratefully acknowledge the support of  a Consolidated Grant (ST/K00926/1) from the UK Science and Technology Funding Council (STFC). PJC and HLG acknowledge support from the European Research Council (ERC-CoG-647939). The computations were performed using Cardiff University's Advanced Research Computing facility (ARCCA).

\bibliographystyle{mn2e}
\bibliography{M31Dust}

\appendix

\section{Converting optical depths to column-densities}\label{APP:convert}

It is common to present images of dust emission in terms of the surface-density of dust, $\Sigma\subD$, or even the associated column-density of hydrogen in all chemical forms, $N_{_{\rm H}}$, because this makes the images easier to conceptualise. If we know the mass opacity coefficient of dust at $300\,\mu{\rm m}$, $\,\kappa_{_{300}}$, then
\begin{eqnarray}
\Sigma\subD&=&\frac{\tau_{_{300}}}{\kappa_{_{300}}}\,.
\end{eqnarray}
If we know the fraction by mass of hydrogen, $X$, and the fraction by mass of dust, $Z\subD$, then
\begin{eqnarray}
N_{_{\rm H}}&=&\frac{X\,\Sigma\subD}{Z\subD\,m_{_{\rm H}}}\;\,=\;\,\frac{X\,\tau_{_{300}}}{Z\subD\,\kappa_{_{300}}\,m_{_{\rm H}}}\,,
\end{eqnarray}
where $m_{_{\rm H}}$ is the mass of an hydrogen atom. The problem is that $Z\subD$, $\kappa_{_{300}}$ and even $X$ are not uniform over the disc of M31. The gas-phase metallicity, $Z$, is observed to decrease by more than an order of magnitude between the centre of M31 and the outer parts; to first order we should assume that $Z\subD$ decreases by a similar factor. Our analysis also indicates that $\beta$ varies, both with galacto-centric radius, and between different environments; these variations are almost certainly accompanied by variations in $\kappa_{_{300}}$. Finally, $X$ probably increases somewhat with galacto-centric radius. Given these sources of uncertainty, and since we do not need $\Sigma\subD$ or $N_{_{\rm H}}$, we work with the far-IR optical
depth, $\tau_{_{300}}$.

\section{Near-IR extinction optical depths from Colour Magnitude Diagrams of Red Giant Branch stars}\label{APP:NIRextinction}

The near-infrared extinction opacity through M31 is estimated using near-infrared colour magnitude diagrams (CMDs) of Red Giant Branch (RGB) stars, and covers a large swathe of M31, comprising approximately one third of the total area, around the major axis on the north-east side of the galaxy, and stretching out to $\sim\!20\,{\rm kpc}$ from the centre \citep{Willetal2014,Dalcetal2015}. This area is divided into $(25\,{\rm pc})^2$ tiles, and the tiles are dithered by $12.5\,{\rm pc}$ to give Nyquist-sampled $25\,{\rm pc}$ resolution. In each tile, Hubble Space Telescope photometry is used to obtain fluxes, ${\cal F}$, in the Wide Field Camera 3/IR $F110W$ and $F160W$ filters, and to construct a CMD of ${\cal F}_{_{F160W}}$ (in the interval 25 to 17 magnitudes) against ${\cal F}_{_{F110}}\!-\!{\cal F}_{_{F160W}}$ (in the interval 0 to 2 magnitudes). This effectively isolates RGB stars, and the analysis takes account of various possible interlopers. The scale-height of RGB stars in M31 is presumed to be much greater ($\ga\!500\,{\rm pc}$) than the scale-height of the dust ($\la\!50\,{\rm pc}$) and the even smaller size of an individual dust cloud ($\la\!10\,{\rm pc}$). Consequently an individual RGB star in M31 is almost certainly either behind, or in front of, most of the dust on its line of sight. Since the intrinsic locus of unreddened RGB stars on the CMD is very narrow, the stars behind the dust layer, and the stars in front of it, end up as distinct populations on the CMD -- unless the reddening is very small -- and hence the optical depth through the dust layer can be estimated. Variations in the intrinsic locus of unreddened RGB stars are handled by constructing reference CMDs from the observed population in regions where (a) the extinction is known to be weak \citep[for example, from dust emission mapping,][]{Draietal2014} and (b) the surface-density of stars is comparable, hence problems due to confusion are similar. Thus the analysis allows for the fact that there are likely to be systematic variations in the intrinsic colours of RGB stars, {\it both} due to the the radial increase in mean stellar age, and the radial decrease in mean stellar metallicity (fortuitously, these two effects tend to cancel each other out), {\it and} across the main star-forming rings at $\sim\!6\,{\rm kpc}$, $\sim\!11\,{\rm kpc}$ and $\sim\!15\,{\rm kpc}$. Variation in the dust optical depth on different lines of sight through the same tile are characterised by a log-normal distribution, with median visual extinction, ${\tilde A}_{_{\rm V}}$ and dimensionless standard deviation, $\sigma$; it is assumed that $A_{_{1.1\mu{\rm m}}}\!=\!0.3266A_{_{\rm V}}$ and $A_{_{1.6\mu{\rm m}}}\!=\!0.2029A_{_{\rm V}}$. By considering a wide range of effects, it is estimated that the resulting optical depths are accurate, except in regions (particularly the outer reaches of M31) where the extinction is low and there are few stars in a given tile, and in regions (near the centre of M31) where the RGB population is very inhomogeneous and there are serious problems with confusion. \\

\section{Dust distributions from detailed modelling}\label{APP:FIRextinction}

The most sophisticated analysis of the dust emission from M31 to date \citep{Draietal2014} combines the six wavelength bands of {\it Herschel} with the seven wavelength bands of {\sc Spitzer}, using a detailed irradiation algorithm \citep{DraineLi2007}. The irradiation algorithm uses a specific dust model, and fits observed fluxes by varying (i) the surface-density of dust, $\Sigma\subD$;  (ii) the fraction of the dust mass that is in PAHs, $q_{_{\rm PAH}}$; (iii) the ambient radiation field, $U_{_{\rm min}}$, which heats most of the dust; (iv) the fraction of dust, $\gamma$, that is more strongly irradiated than $U_{_{\rm min}}$; and (v) the fraction of the dust mass that is very strongly irradiated (in PDRs). The dust model allows for a distribution of grain compositions and sizes, and for PAHs to be transiently heated; as with \pp, there is a distribution of temperatures along each line of sight. In a post-processing step, a notional equilibrium dust temperature, ${\tilde T}$, is derived on the basis of the mean radiation intensity, and a notional emissivity index, ${\tilde\beta}$, is estimated from the mismatch between the observed and modelled fluxes at $250\,\mu{\rm m}$ and $500\,\mu{\rm m}$.

\section{Theoretical dust models}\label{APP:DustModels}

Table \ref{TAB:DustModels1} gives values of $\beta$ (Column 1) and ${\cal R}^{\mbox{\tiny model}}_\kappa$ (Column 2) for commonly used dust models from the literature, along with a brief indication of the ingredients of the model (Column 3), and the source reference (Column 4). These are the models plotted on Fig. \ref{FIG:calR-beta}; they are also plotted on Fig. \ref{FIG:calR-beta_X}. ${\cal R}^{\mbox{\tiny model}}_\kappa$ is computed on the assumption that, when convolved with an average RGB spectrum, the mean wavelength of the F110W filter is $1.14\,\mu{\rm m}$.

Table \ref{TAB:DustModels2} gives the same information for dust models computed using Mie Theory and optical constants from the literature. In all these models we assume a single grain radius $r\subD$, and in all but one case we adopt $r\subD\!=\!0.1\,\mu{\rm m}$; the exception is model 9a where we adopt $r\subD\!=\!0.01\,\mu{\rm m}$. These models are plotted on Fig. \ref{FIG:calR-beta_X}, unless they fall outside its boundaries, i.e. ${\bar\beta}$ outside the range $[1.00,3.00]$ or ${\cal R}^{\mbox{\tiny model}}_\kappa$ outside the range $[0,10^4]$; this excludes fourteen models. We see that many of the single-size models are  clustered round $({\bar\beta},{\cal R}^{\mbox{\tiny model}}_\kappa)=(2.0,400)$. Only four models populate the region of high ${\bar\beta}$ and low ${\cal R}^{\mbox{\tiny obs.}}_\tau$ observed in the star-forming rings of M31; these are models 10, 13d, 13e and 13j.

\begin{table*}\begin{center}
\caption{Single-size models computed using optical constants from the literature. Columns 1 and 2 give the values of $\beta$ and ${\cal R}^{\mbox{\tiny model}}_\kappa$. Columns 3 and 4 give the mineralogy and the source reference. Column 5 gives the ID used to identify these models on Fig. \ref{FIG:calR-beta_X}. Values of $\beta$ and ${\cal R}^{\mbox{\tiny model}}_\kappa$ that populate the high $\beta$ and low ${\cal R}^{\mbox{\tiny model}}_\kappa$ area of the plot (10,13d, 13e and 13j) are in bold, as are their IDs. The IDs of models that fall outside Fig. \ref{FIG:calR-beta_X} are in itallics}
\begin{tabular}{crclll}\hline
$\hspace{1.0cm}\beta\hspace{0.9cm}$ & ${\cal R}\;\;$ & \hspace{0.5cm} & {\sc Model Mineralogy}\hspace{2.3cm} & {\sc Source}\hspace{4.9cm} & {\sc ID} \\\hline
2.04 & 5090 & \hspace{0.5cm} & graphite-parallel & \citet{DrainLee1984} & 5a \\
2.01 & 334 & \hspace{0.5cm} & graphite-perpendicular & \citet{DrainLee1984} & 5b \\
2.02 & 560 & \hspace{0.5cm} & astronomical silicate & \citet{DrainLee1984} & 5c \\
1.51 & 3870 & \hspace{0.5cm} & silicon carbide & \citet{Pegourie1988} & 6 \\
0.98 & 826 & \hspace{0.5cm} & amC(AC1) & \citet{RoulMart1991} & {\it 7a} \\
1.57 & 2250 & \hspace{0.5cm} & benzene & \citet{RoulMart1991} & 7b \\
2.00 & 1170 & \hspace{0.5cm} & circumstellar O-poor silicate & \citet{Osseetal1992} & 8a \\
2.00 & 1370 & \hspace{0.5cm} & circumstellar O-rich silicate & \citet{Osseetal1992} & 8b \\
2.00 & 1180 & \hspace{0.5cm} & interstellar O-poor silicate & \citet{Osseetal1992} & 8c \\
2.00 & 1390 & \hspace{0.5cm} & interstellar O-rich silicate & \citet{Osseetal1992} & 8d \\
1.92 & 1400 & \hspace{0.5cm} & neutral PAH & \citet{LaorDrai1993} & 9a \\
2.00 & 13600 & \hspace{0.5cm} & silicon carbide & \citet{LaorDrai1993} & {\it 9b} \\
{\bf 2.19} & {\bf 356} & \hspace{0.5cm} & cosmic silicate & \citet{Jaegetal1994} & {\bf 10} \\
0.74 & 792 & \hspace{0.5cm} & oxide, Mg:Fe=60:40 & \citet{Hennetal1995} & {\it 11a} \\
0.66 & 1120 & \hspace{0.5cm} & oxide, Mg:Fe=50:50 & \citet{Hennetal1995} & {\it 11b} \\
0.64 & 1040 & \hspace{0.5cm} & oxide, Mg:Fe=30:70 & \citet{Hennetal1995} & {\it 11c} \\
0.67 & 1040 & \hspace{0.5cm} & oxide, Mg:Fe=20:80 & \citet{Hennetal1995} & {\it 11d} \\
0.63 & 994 & \hspace{0.5cm} & oxide, Mg:Fe=10:90 & \citet{Hennetal1995} & {\it 11e} \\
0.87 & 1030 & \hspace{0.5cm} & oxide, Mg:Fe=0:100 & \citet{Hennetal1995} & {\it 11f} \\
2.00 & 260 & \hspace{0.5cm} & enstatite & \cite{Dorsetal1995} & 12a \\
1.97 & 270 & \hspace{0.5cm} & pyroxene, Mg:Fe=95:5 & \cite{Dorsetal1995} & 12b \\
1.99 & 286 & \hspace{0.5cm} & pyroxene, Mg:Fe=80:20 & \cite{Dorsetal1995} & 12c \\
1.99 & 321 & \hspace{0.5cm} & pyroxene, Mg:Fe=70:30 & \cite{Dorsetal1995} & 12d \\
1.99 & 372 & \hspace{0.5cm} & pyroxene, Mg:Fe=60:40 & \cite{Dorsetal1995} & 12e \\
1.99 & 452 & \hspace{0.5cm} & pyroxene, Mg:Fe=50:50 & \cite{Dorsetal1995} & 12f \\
1.99 & 516 & \hspace{0.5cm} & pyroxene, Mg:Fe=40:60, $\,0.1\,\mu{\rm m}$ & \cite{Dorsetal1995} & 12g \\
1.99 & 1220 & \hspace{0.5cm} & olivine & \cite{Dorsetal1995} & 12h \\
1.99 & 1220 & \hspace{0.5cm} & glassy olivine & \cite{Dorsetal1995} & 12i \\
1.99 & 452 & \hspace{0.5cm} & glassy pyroxene & \cite{Dorsetal1995} & 12j \\
2.02 & 319 & \hspace{0.5cm} & olivine, Mg:Fe=100:0 & \citet{HennStog1996} & 13a \\
2.01 & 734 & \hspace{0.5cm} & olivine, Mg:Fe=70:30 & \citet{HennStog1996} & 13b \\
1.97 & 964 & \hspace{0.5cm} & olivine, Mg:Fe=60:40 & \citet{HennStog1996} & 13c \\
{\bf 2.63} & {\bf 447} & \hspace{0.5cm} & orthopyroxene, Mg:Fe=100:0 & \citet{HennStog1996} & {\bf 13d} \\
{\bf 2.42} & {\bf 353} & \hspace{0.5cm} & orthopyroxene, Mg:Fe=70:30 & \citet{HennStog1996} & {\bf 13e} \\
1.97 & 428 & \hspace{0.5cm} & orthopyroxene, Mg:Fe=60:40 & \citet{HennStog1996} & 13f \\
1.78 & 5990 & \hspace{0.5cm} & iron & \citet{HennStog1996} & 13g \\
0.43 & 7620 & \hspace{0.5cm} & troilite & \citet{HennStog1996} & {\it 13h} \\
2.01 & 431 & \hspace{0.5cm} & organics & \citet{HennStog1996} & 13i \\
{\bf 3.89} & {\bf 235} & \hspace{0.5cm} & water ice & \citet{HennStog1996} & {\bf 13j} \\
1.57 & 5970 & \hspace{0.5cm} & a-C(BE) & \citet{Zubketal1996} & 14a \\
1.46 & 2530 & \hspace{0.5cm} & a-C(ACAR) & \citet{Zubketal1996} & 14b \\
2.05 & 1870 & \hspace{0.5cm} & a-C(ACH2) & \citet{Zubketal1996} & 14c \\
1.90 & 6350 & \hspace{0.5cm} & a-C & \citet{Hannetal1998} & 15 \\
1.21 & 3380 & \hspace{0.5cm} & cellulose, $400\,{\rm K}$ & \citet{Jaegetal1998} & 16a \\
1.44 & 23900 & \hspace{0.5cm} & cellulose, $600\,{\rm K}$ & \citet{Jaegetal1998} & {\it 16b} \\
1.99 & 2520 & \hspace{0.5cm} & cellulose, $800\,{\rm K}$ & \citet{Jaegetal1998} & 16c \\
2.09 & 4250 & \hspace{0.5cm} & cellulose, $1000\,{\rm K}$ & \citet{Jaegetal1998} & 16d \\
2.03 & 9600 & \hspace{0.5cm} & crystalline olivine & \citet{Fabietal2001} & 17a \\
2.04 & 14700 & \hspace{0.5cm} & crystalline fayalite & \citet{Fabietal2001} & {\it 17b} \\
2.01 & 11200 & \hspace{0.5cm} & spinel & \citet{Fabietal2001} & {\it 17c} \\
2.01 & 7690 & \hspace{0.5cm} & spinel, $950\,\rm{^oC}$ & \citet{Fabietal2001} & 17d \\
1.90 & 354 & \hspace{0.5cm} & enstatite & \citet{Jaegetal2003b} & 18a \\
1.98 & 362 & \hspace{0.5cm} & forsterite & \citet{Jaegetal2003b} & 18b \\
2.09 & 5230 & \hspace{0.5cm} & perovskite & \citet{Poscetal2003} & 19a \\
2.02 & 8560 & \hspace{0.5cm} & anatase & \citet{Poscetal2003} & 19b \\
2.02 & 34000 & \hspace{0.5cm} & brookite & \citet{Poscetal2003} & {\it 19c} \\
1.31 & 3950 & \hspace{0.5cm} & a-C & \citet{JonesAP2012} & 20a \\
1.16 & 2020 & \hspace{0.5cm} & a-C(:H) & \citet{JonesAP2012} & 20b \\
2.05 & 4620 & \hspace{0.5cm} & a-Sil (Mg-rich pyroxene) & \citet{Koehetal2014} & 21a \\
1.99 & 3680 & \hspace{0.5cm} & a-Sil (Mg-rich olivine) & \citet{Koehetal2014} & 21b \\
1.98 & 8030 & \hspace{0.5cm} & magnetite & Triaud, unpublished & 22a \\
2.03 & 75100 & \hspace{0.5cm} & hematite & Triaud, unpublished & {\it 22b} \\\hline
\end{tabular}\label{TAB:DustModels2}
\end{center}\end{table*}

\label{lastpage}
\end{document}